\documentclass[sigconf]{acmart}

\AtBeginDocument{%
  }

\setcopyright{acmlicensed}
\copyrightyear{2025}
\acmYear{2025}
\acmDOI{XXXXXXX.XXXXXXX}

\acmConference[IUI'25]{30th Annual ACM Conference on Intelligent User Interfaces}{March 24--27,
  2025}{Cagliari, Italy}
\acmISBN{978-1-4503-XXXX-X/18/06}

\usepackage{listings}
\usepackage{color} 

\usepackage{array}
\usepackage{colortbl}
\usepackage{xcolor}
\usepackage{booktabs}
\usepackage{tabularx}

\usepackage{siunitx} 
\usepackage{verbatim}

\usepackage{courier}  
\usepackage{setspace}

\definecolor{lightgray}{RGB}{245,245,245}
\definecolor{orange}{RGB}{240,148,0}
\definecolor{codeblue}{RGB}{100,149,237}
\definecolor{codered}{RGB}{205,92,92}

\definecolor{lightbg}{RGB}{248,250,252}

\definecolor{codegreen}{rgb}{0,0.6,0}
\definecolor{codegray}{rgb}{0.5,0.5,0.5}
\definecolor{codepurple}{rgb}{0.58,0,0.82}

\lstdefinestyle{mystyle}{
    backgroundcolor=\color{backcolour},
    commentstyle=\color{codegreen},
    keywordstyle=\color{magenta},
    numberstyle=\tiny\color{codegray},
    stringstyle=\color{codepurple},
    basicstyle=\ttfamily\footnotesize,
    breakatwhitespace=false,
    breaklines=true,
    captionpos=b,
    keepspaces=true,
    numbers=left,
    numbersep=5pt,
    showspaces=false,
    showstringspaces=false,
    showtabs=false,
    tabsize=2
}

\begin{document}


\title{Text-to-SQL Domain Adaptation via Human-LLM Collaborative Data Annotation}

%

\author{Yuan Tian}
\authornote{This work was done during the first author's internship at Adobe.}
\email{tian211@purdue.edu}
\affiliation{%
  \institution{Purdue University}
  \city{West Lafayette}
  \state{Indiana}
  \country{USA}
}

\author{Daniel Lee}
\email{dlee1@adobe.com}
\affiliation{%
  \institution{Adobe Inc.}
  \city{San Jose}
  \state{California}
  \country{USA}
}

\author{Fei Wu}
\email{feiw@adobe.com}
\affiliation{%
  \institution{Adobe Inc.}
  \city{Seattle}
  \state{Washington}
  \country{USA}
}

\author{Tung Mai}
\email{tumai@adobe.com}
\affiliation{%
  \institution{Adobe Inc.}
  \city{San Jose}
  \state{California}
  \country{USA}
}

\author{Kun Qian}
\email{kunq@adobe.com}
\affiliation{%
  \institution{Adobe Inc.}
  \city{Seattle}
  \state{Washington}
  \country{USA}
}

\author{Siddhartha Sahai}
\email{siddharthas@adobe.com}
\affiliation{%
  \institution{Adobe Inc.}
  \city{Seattle}
  \state{Washington}
  \country{USA}
}

\author{Tianyi Zhang}
\email{tianyi@purdue.edu}
\affiliation{%
  \institution{Purdue University}
  \city{West Lafayette}
  \state{Indiana}
  \country{USA}
}

\author{Yunyao Li}
\email{yunyaol@adobe.com}
\affiliation{%
  \institution{Adobe Inc.}
  \city{San Jose}
  \state{California}
  \country{USA}
}


\newcommand{\tool}{\textsc{SQLsynth}}
\newcommand{\todo}[1]{\textcolor{red}{#1}}
\newcommand{\circled}[1]{{\large \textcircled{\footnotesize #1}}}

\newcommand{\edit}[1]{#1}

\begin{abstract}
    Text-to-SQL models, which parse natural language (NL) questions to executable SQL queries, are increasingly adopted in real-world applications. 
    However, deploying such models in the real world often requires adapting them to the highly specialized database schemas used in specific applications. 
    We find that existing text-to-SQL models experience significant performance drops when applied to new schemas, primarily due to the lack of domain-specific data for fine-tuning. This data scarcity also limits the ability to effectively evaluate model performance in new domains.
    Continuously obtaining high-quality text-to-SQL data for evolving schemas is prohibitively expensive in real-world scenarios.   
    To bridge this gap, we propose {\tool}, a human-in-the-loop text-to-SQL data annotation system.
    {\tool} streamlines the creation of high-quality text-to-SQL datasets through human-LLM collaboration in a structured workflow.
    A within-subjects user study comparing {\tool} with manual annotation and ChatGPT shows that {\tool} significantly accelerates text-to-SQL data annotation, reduces cognitive load, and produces datasets that are more accurate, natural, and diverse. \edit{Our code is available at \url{https://github.com/magic-YuanTian/SQLsynth}\footnote{Backup for the official repo \url{https://github.com/adobe/nl_sql_analyzer}}.}
\end{abstract}


\begin{teaserfigure}
  \centering
  \includegraphics[width=\textwidth]{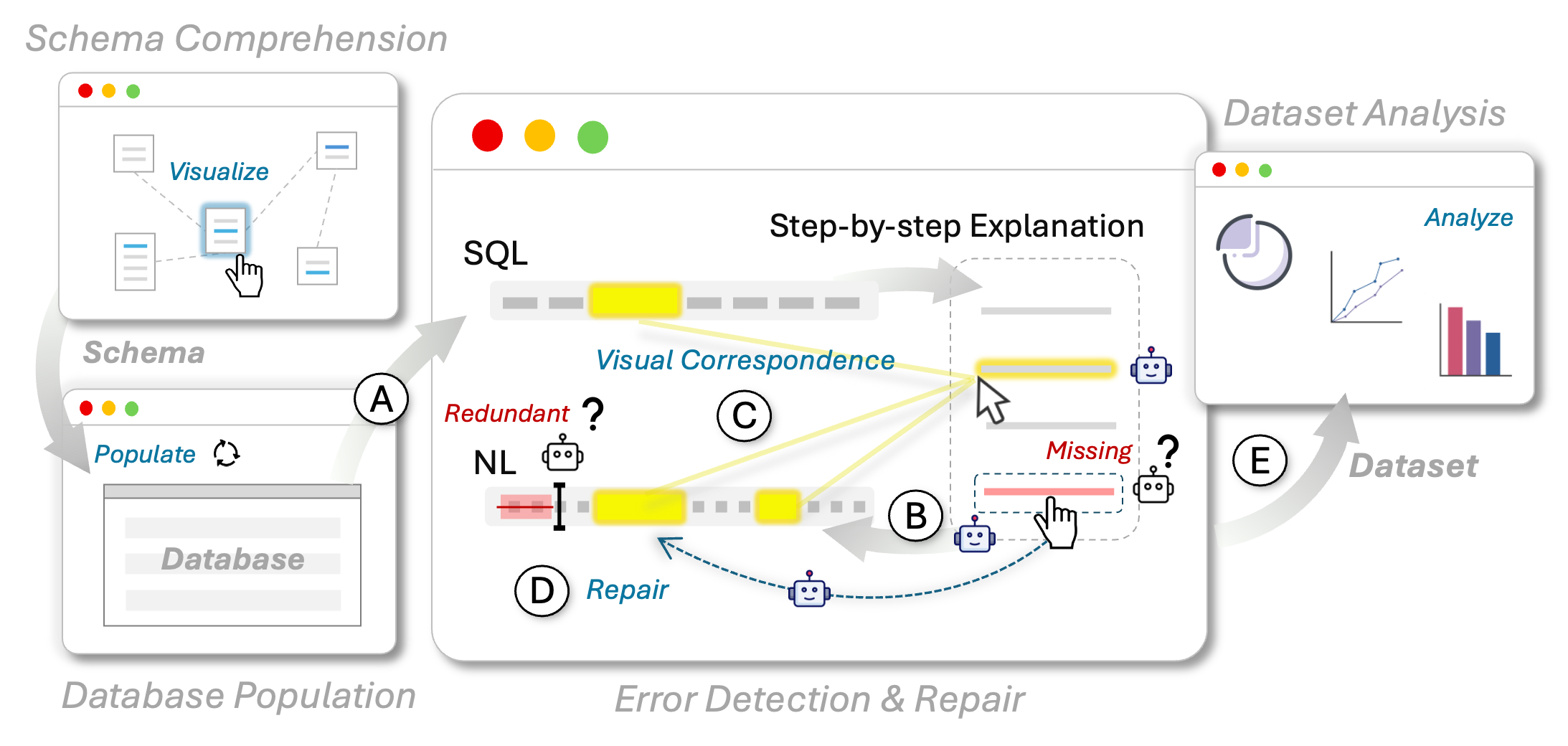}
  \caption{An overview of {\tool}: \textbf{(A)} Given a database schema, {\tool} populates a database and sample SQL queries for the database. \textbf{(B)} After a step-by-step analysis of each sampled SQL query, {\tool} translates the query into a natural language (NL) question. \textbf{(C)} The large language model (LLM) aligns the NL question with the SQL query through the step-by-step analysis. Users can hover over each step to check the corresponding span in both SQL and NL. \textbf{(D)} {\tool} detects potential errors in the NL and highlights them in red. Users can repair the NL by injecting missing steps or deleting redundant text. \textbf{(E)} Users monitor and visualize the composition of the annotated dataset, thereby controlling the annotation process.
    }
  \label{fig:teaser}
\end{teaserfigure}


\maketitle

\section{Introduction}

Natural language interfaces to databases can significantly democratize complex data analysis and decision-making processes for end-users by allowing them to express their intent through NL.
These interfaces are mainly powered by text-to-SQL models that convert NL questions into SQL queries~\cite{smbop, picard, dinsql}. The resulting SQL queries are then executed on the database to retrieve the results for users.
Recent advancements, notably in large language models (LLMs), have substantially enhanced NL interfaces and opened a promising market that is expected to grow around three times in five years~\cite{market1, market2, market3}.


Despite these advancements, the accuracy and reliability of text-to-SQL models in real-world applications remain suboptimal, especially in high-stakes domains such as finance and healthcare where an error can cause severe consequences. 
One root cause of this inaccuracy is the domain shift, which is a common challenge in machine learning~\cite{domain_adaption1, domain_adaption2, domain_adaption3}. 
Specifically, deploying an NL interface to a real-world database requires the underlying text-to-SQL model to understand the database schema, which is often highly specialized with unique architectures and data contexts that public datasets such as Spider~\cite{spider} and BIRD~\cite{bird} do not cover. 
Our formative study with engineers from Adobe reveals a text-to-SQL accuracy drop of at least 13.3\% for newly added columns and 9.1\% for new tables. Moreover, as the database schema evolves over time, the performance of the text-to-SQL model continues to decline. To fully unleash the computational power of text-to-SQL models in the new domain, there is a demand for annotating sufficient pairs of text and SQL data under a target schema for fine-tuning a text-to-SQL model to the domain. 


Recent studies have shown that  LLMs can be used as data annotators in domains such as image captioning, code generation, and topic detection~\cite{llm_data_annotation, annollm, llm_annotator_1, chatgpt_outperform_crowd_workers, self_instruct, oss_instruct, evol_instruct}.
While we could use LLMs to annotate text-to-SQL datasets, it introduces new challenges such as hallucination~\cite{llm_hallucination}. 
Furthermore, LLMs are inherently not designed or trained to generate diverse data.
They frequently generate repetitive data and struggle to maintain consistency in large datasets due to the limited attention in a long context~\cite{SPA, longformer}. Therefore, we believe that when harnessing LLMs for curating text-to-SQL datasets, it is essential to incorporate human knowledge and improve human control. To the best of our knowledge, no prior work has investigated human-in-the-loop data annotation methods in the domain of text-to-SQL generation.


\begin{figure*}[ht]
  \centering
  \includegraphics[width=0.8\textwidth]{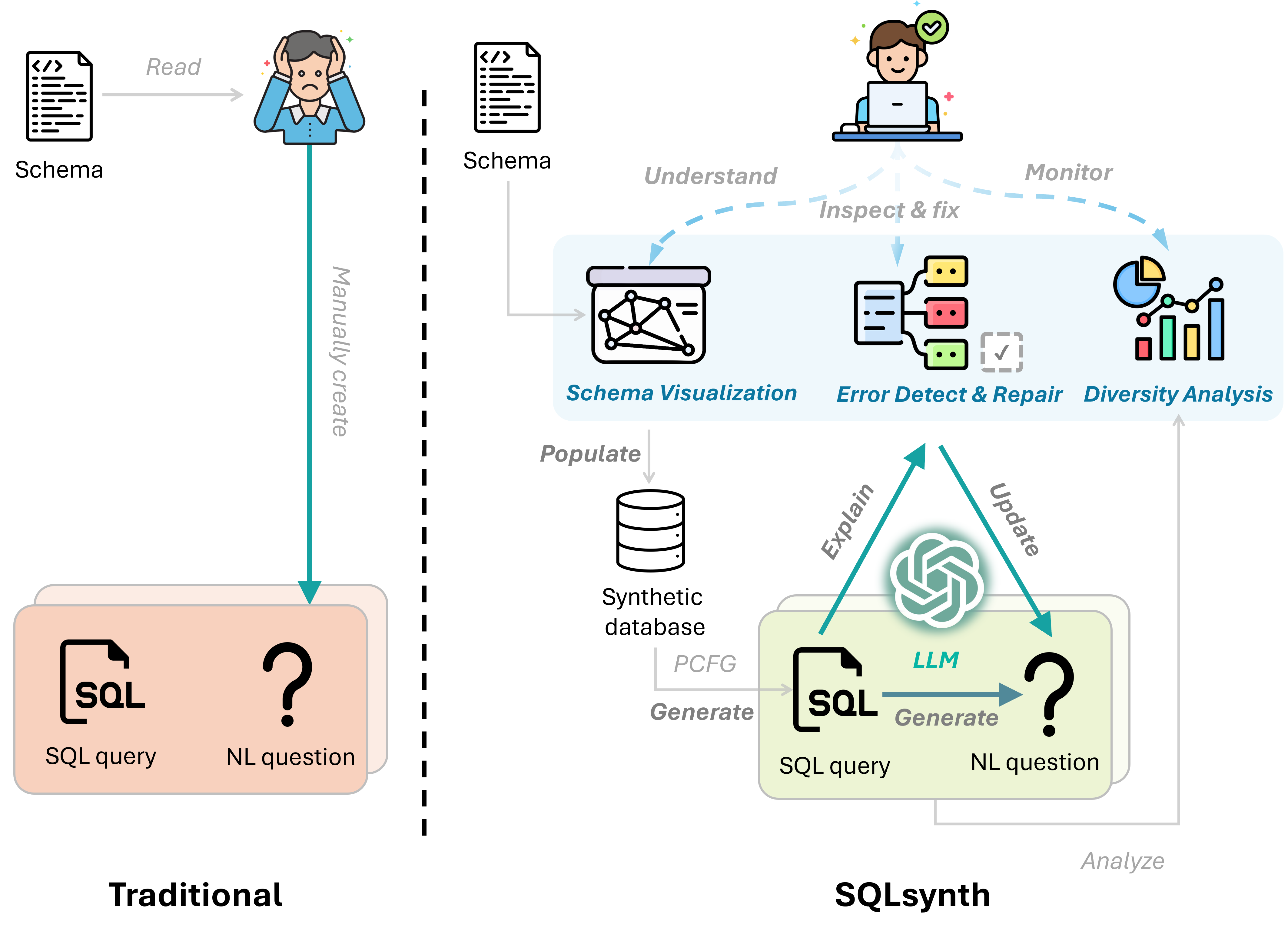}
  \caption{Illustration of traditional text-to-SQL annotation v.s. using {\tool}. \edit{Green arrows indicate dataset creation steps, gray arrows represent supportive data flows, and blue dashed arrows show user interactions with the interface. {\tool} includes three interactive features in the blue box: {\em schema visualization}, {\em error detection and repair}, and {\em diversity analysis}. Annotators can leverage these features to efficiently control the data annotation process.}}
  \label{fig:overview}
\end{figure*}

To bridge this gap, we propose {\tool}, an interactive text-to-SQL data annotation system designed to streamline the process of creating high-quality text-to-SQL datasets on a customized schema. 
Our insight is to automate labor-intensive procedures (e.g., initialize SQL queries and NL questions) while providing effective error detection and repair mechanisms, allowing annotators to efficiently verify and refine the data in collaboration with the LLM.
Figure~\ref{fig:overview} compares the pipeline of traditional text-to-SQL annotation with the pipeline when using {\tool}. 
Traditionally, given a database schema, annotators have to understand the schema architecture and manually create new SQL queries and NL questions from scratch. This process is often mentally demanding and prone to biases, as personal assumptions can affect query patterns and phrasing.

In contrast, {\tool} visualizes the database schema as an interactive editable graph to facilitate comprehension. Based on the customized schema, {\tool} leverages a rule-based method to populate a database with numerous diverse records. Then, it uses this populated database to sample SQL queries using Probabilistic Context-Free Grammar (PCFG).
{\tool} leverages both a grammar-based approach and an LLM to analyze SQL queries step by step. Based on the analysis, it then translates the SQL query into a corresponding NL question. The LLM automatically aligns the NL question with the SQL query through this step-by-step analysis. Based on this alignment, {\tool} identifies potential errors and allows annotators to efficiently rectify these issues by instructing the LLM through a structured workflow.
Specifically, annotators can inject missing information from misaligned steps or delete redundant text in the NL question.
Finally, {\tool} analyzes the annotated data, providing dynamic dataset composition on property distributions such as statistics for specific referenced entities in the current dataset. This feature enables annotators to iteratively monitor and refine the dataset's diversity and quality.

We evaluated {\tool} through a within-subjects study with 12 participants. Compared to manual annotation and using ChatGPT, participants using {\tool} annotated 4.4X and 2.3X more data,  with 87\% and 84\% fewer errors, respectively.
We evaluated the naturalness of annotated data using both the Flesch-Kincaid Score and human ratings. The results indicate that NL questions annotated by {\tool} are more natural compared to those produced by ChatGPT.
To evaluate dataset diversity, we analyze the number of clauses, tables, columns, and values in SQL queries. The results show that {\tool} achieves a higher Simpson's Diversity Index, suggesting better overall diversity.
Furthermore, during the annotation, participants self-reported experiencing less cognitive load and greater confidence when using {\tool}.

In summary, this work makes the following contributions:

\begin{itemize}
    \item A \textbf{formative study with industrial practitioners} identifies five major users' needs of text-to-SQL annotation in practice.
    
    \item A \textbf{comprehensive interactive text-to-SQL annotation system}, {\tool}, streamlines the annotation of text-to-SQL datasets. It facilitates database comprehension, database population, text-to-SQL data generation, error detection and repair, and dataset analysis.
    
    
    \item A \textbf{comprehensive evaluation} assesses the usability and effectiveness of {\tool}, demonstrating significant improvements in annotation speed, accuracy, naturalness, and dataset diversity compared to both manual efforts and conversational AI assistants like ChatGPT.
\end{itemize}

\section{Related Workd}

\subsection{SQL Generation from Natural Language}


Developing a natural language interface for databases has been a long-standing problem since the 1970s. In 1972,  LUNAR~\cite{lunar} was proposed to enable geologists to query the chemical analysis data of lunar rocks. 
Early effort in this domain focuses on logic-based~\cite{logic1, logic2} and rule-based~\cite{rule1, SQLizer, ATHENA, Semantic-Tractability, construct_interface} approaches. However, these approaches require significant human efforts to create the translation rules and are limited to a definite set of queries~\cite{where-are-we, Semantic-Tractability}.

The emergence of many public text-to-SQL datasets~\cite{spider, wikisql, SQLizer, bird} propel researchers to develop learning-based text-to-SQL models~\cite{smbop, STEPS, ratsql, editsql, picard, sqlova}.
Early text-to-SQL models mainly leverage the encoder-decoder architecture~\cite{lin2020bridging, zhong-etal-2020-grounded, seq2sql}, where the encoder maps NL question and database schema to latent state and decoder maps the latent state to a SQL query. 
Recently, more work~\cite{xiyan, dailsql, distillery} has focused on leveraging large language models (LLMs).
These LLMs are pre-trained on large corpora and exhibit general ability in semantic parsing. 
They can effectively solve text-to-SQL tasks in a zero-shot setting with prompts or a few demonstrations~\cite{thorpe2024dubosqldiverseretrievalaugmentedgeneration, -dialsql}.
In the meanwhile, some works fine-tune existing LLMs to achieve better performance LLMs~\cite{codes, finetune_sql_llm, sqlcoder, NSText2SQL}. However, these approaches are optimized for general text-to-SQL capabilities and may not perform well on specialized schemas. {\tool} bridges this gap by enabling instant text-to-SQL dataset acquisition for a specific schema, thereby facilitating focused fine-tuning and evaluation.

\subsection{Text-to-SQL Dataset Creation}



\subsubsection{Human-annotated Text-to-SQL Datasets}
There has been a growing number of text-to-SQL datasets~\cite{atis, scholar, restaurants, geoquery, academic, yelp_and_IMDB, advising, wikisql, spider, spider_syn, spider_dk, dusql, kaggledbqa, bird}. 
For example, WikiSQL~\cite{wikisql} contains 80,654 hand-annotated examples of questions and SQL queries across 24,241 Wikipedia tables. Spider~\cite{spider} is a cross-domain text-to-SQL dataset annotated by 11 college students. It contains 10,181 questions and 5,693 SQL queries across 200 databases, spanning 138 different domains. Another notable dataset, BIRD~\cite{bird}, includes 12,751 queries over 95 databases spanning 37 professional domains created by crowd-sourcing.

While these datasets attempt to incorporate real-world query scenarios, they are limited by the number of schemas. In practice, different applications may use significantly different schema architectures and query traffic, leading to significant disparities compared to existing datasets. Moreover, while each dataset contains thousands of queries in total, the number of queries per database remains insufficient to cover all query types, due to the high cost of human annotation.





\subsubsection{Text-to-SQL Data Augmentation \& Synthesis} Compared with manual annotation, there have been a lot of efforts in automating the dataset construction process. The first line of research focuses on data augmentation from existing datasets, which transforms or expands an existing dataset to a new dataset~\cite{syntaxsqlnet, conda, nlp_synthetic_1, nlp_synthetic_2, nlp_synthetic_3, nlp_synthetic_4}.
For example, SyntaxSQLNet~\cite{syntaxsqlnet} generates cross-domain data using query templates from the Spider dataset. CONDA~\cite{conda} augments SParC~\cite{sparc} and CoSQL~\cite{cosql} by modifying SQL dialogue states to generate different queries and questions.
However, such data augmentation is limited to the existing dataset, making it challenging to significantly increase the diversity of the data.

To address the diversity limitation of data augmentation, the second line of research focuses on sampling various queries based on the SQL grammar and then translating them into NL questions~\cite{conda, simiar_algorithm}. 
While grammar-based sampling is feasible, translating SQL queries back to NL questions remains a challenge. Existing methods are either template-based~\cite{STEPS, sqlucid, logos, explaininnl, SQL-to-text, diy, NaLIR} or model-based~\cite{IRnet, sql_to_text2, sql_to_text3, PCFG_SQL_synthesize}, both of which have limitations. Template-based methods translate formal queries into NL questions based on pre-defined templates, which lack diversity and naturalness in generated questions. Model-based methods train or employ a pre-trained model to generate the NL question from the formal query. 
For example, \cite{PCFG_SQL_synthesize} trained a BART-based~\cite{bart} translation model that maps a SQL to an NL question. 
However, such translation models can introduce errors and make the dataset unreliable.

Recently, research has demonstrated that LLMs can work as effective data annotators~\cite{llm_data_annotation, annollm, llm_annotator_1, chatgpt_outperform_crowd_workers, self_instruct, oss_instruct, evol_instruct}. 
While LLMs can be prompted to generate text-to-SQL data, they are not well-suited for producing large amounts of diverse data from scratch. 

To address these issues, {\tool} first samples a SQL query based on grammar rules and translates it into an NL question using an LLM. To handle potential translation errors, {\tool} adopts a human-in-the-loop inspection and repair method. It detects and highlights any potential misalignment between the SQL query and NL questions for user inspection. It then allows users to fix the alignment issue by injecting missing information and removing redundant information in collaboration with the LLM.




\subsection{Interactive Data Annotation}

While pure synthetic datasets are not reliable, human annotation is time-consuming and labor-intensive.
To strike a balance, interactive data annotation methods~\cite{semi_automatic_data_annotation, ARAIDA, INCEpTION, INCEpTION2, interactive_video_object_mask_annotation, freeal, fitannotator} were introduced to reduce annotation effort while maintaining annotation quality. 
These approaches typically employ an annotation model that provides suggestions for human reviewers to approve or correct.

Compared to manual annotation, interactive annotation offers a significant advantage in that human annotators need only verify model-generated annotations rather than creating data from scratch. This can significantly accelerate the annotation process while reducing human effort.
A common strategy of interactive annotation is utilizing active learning \cite{active_learning1, active_learning2, active_learning3, fitannotator, INCEpTION}, which strategically selects the most informative data points for annotation. This approach aims to optimize the model's performance incrementally with the least amount of human-annotated data.
However, a major limitation of active learning is its tendency to reinforce data biases. Although this approach selectively samples data points believed to be most valuable for model improvement, it may inadvertently focus on atypical examples that do not represent the full spectrum of the dataset. Consequently, the model trained on such datasets may develop a skewed understanding, resulting in poor performance.

Facilitating efficient collaboration between intelligent systems and humans has long been a central theme in HCI research, initially introduced in the seminal work on man-computer symbiosis~\cite{man-computer}. Nowadays, the imperfections of AI models in high-stakes domains underscore the need for enhanced human-AI collaboration. Interactive data annotation exemplifies this type of collaboration, aiming for more accurate and trustworthy outcomes. However, to the best of our knowledge, no effective interactive text-to-SQL annotation tools existed prior to this work.

\section{Formative Study}

To understand the specific requirements for text-to-SQL dataset annotation, we conducted a formative study by interviewing 5 engineers from Adobe. These interviewees have experienced annotating text-to-SQL datasets in their work.
We describe our interview process in Section~\ref{sec:interview}. Based on these interviews, we identified five major user needs in Section~\ref{sec:user_needs}. 
Finally, we discuss our design rationale in Section~\ref{sec:design}, aiming to address the user needs.

\subsection{Interview}
\label{sec:interview}

We conducted 20-minute semi-structured interviews with each interviewee through a conversational and think-aloud process. 
During these interviews, we first asked about the \textbf{motivation} for text-to-SQL annotation in their use cases, specifically about the schemas they worked on and why obtaining more data was important.
Interviewees reported that when deploying a new service, companies often needed to introduce new entities and restructure the original schema.
However, after updating the schema, they typically found that model performance dropped dramatically. Their regression tests showed an overall accuracy drop of 13.3\% for newly added columns and 9.1\% for new tables. As the schema was further updated, performance continued to decline. Moreover, as the schema changed significantly, they needed a large amount of new data on the updated schema to ensure a robust evaluation.

Second, we asked about their detailed \textbf{workflow} and whether they used any tools to assist with data annotation. Interviewees reported that they did not use any specific tool for annotation, although they sometimes asked ChatGPT to generate initial data. 
Additionally, they often leveraged previous datasets by adapting previous queries to the new schema, such as replacing an outdated column name with a new one.
After annotation, their colleagues performed peer reviews to check and refine the data.

Third, we asked about \textbf{challenges} they had met and the speed of their dataset annotation. Overall, they considered annotation to be very expensive. 
Interviewees mentioned that one engineer could only annotate 50 effective SQL and NL pairs per day in their use case. 
They often lost track and felt overwhelmed during annotation. 
Despite the peer review, they still felt a lack of confidence in the quality of the annotated data. 
They pointed out that randomness existed throughout the entire procedure. 
We summarize more challenges as user needs in Section~\ref{sec:user_needs}.

\subsection{User Needs}
\label{sec:user_needs}

\noindent \textbf{\textit{N1: Effective Schema Comprehension.}} 
Text-to-SQL annotation assumes that users can easily understand the database schema specified in a certain format (e.g., Data Definition Language). However, our interviews indicate that it is cumbersome and error-prone for users to navigate and comprehend complex schemas from such a specification format.

\noindent \textbf{\textit{N2: Creating New Queries.}}
Creating SQL queries requires a deep understanding of both database schema and SQL grammar. When creating a text-to-SQL dataset, users need to continually come up with new, diverse SQL queries. However, it is challenging for them to break free from preconceptions shaped by existing queries they have seen before.

\noindent \textbf{\textit{N3: Detecting Errors in the Annotated Data.}} 
An annotated dataset may include errors, which can deteriorate model performance and evaluation results.
Our interviews suggest that annotators need an effective mechanism for detecting potential errors or ambiguity in the constructed queries.

\noindent \textbf{\textit{N4: Efficiently Correcting the Detected Errors.}} After identifying errors, users need an efficient way to correct these errors to ensure the accuracy and reliability of the dataset. They need to ensure the SQL query is syntactically correct, and the NL is semantically equivalent to the SQL query.

\noindent \textbf{\textit{N5: Improve Dataset Diversity.}}
Dataset diversity is crucial for improving model performance and ensuring rigorous evaluation.
Human annotation can easily introduce biases due to individual knowledge gaps and a lack of holistic understanding of the dataset composition. 
Thus, interviewees reported the need for an effective way to improve diversity and eliminate biases in the dataset.

\subsection{Design Rationale}
\label{sec:design}

To support \textbf{N1}, our approach visualizes the database schema as a dynamic, editable graph. This enables users to quickly grasp the overall structure of the database and the relationships between entities. Users can explore detailed information such as data type through further interactions with the graph.

To support \textbf{N2}, our approach alleviates the burden of manually creating new SQL queries. We design an algorithm to randomly sample SQL query templates based on SQL grammar, then fill out this template with entities and values retrieved from the database. We make the SQL generation highly configurable---users can manually adjust keyword probability, or automatically tune the probability by an existing dataset.

To support \textbf{N3}, our approach renders the alignment between the SQL query and the NL question via a step-by-step analysis. Our approach then prompts the LLM to highlight potential misalignments to users.
Subsequently, our approach performs a textual analysis to check the equivalence of the SQL query and NL question and offers users a confidence score about their consistency.

To support \textbf{N4}, we handle two common errors---missing information and including irrelevant information in the NL question. Our approach allows users to fix errors by injecting missing information or removing irrelevant details based on LLM-generated suggestions.

To support \textbf{N5}, our approach first enables users to sample SQL queries based on a probability distribution learned from real-world data rather than creating them manually. 
Furthermore, our approach supports visualizing various dataset compositions through diagrams. 
For example, users can view a bar chart displaying the distribution of column counts in SQL queries. This feature allows users to monitor dataset composition during annotation, maintaining control over the annotation direction and improving data diversity.

\section{Implementation}

{\tool} comprises multiple UI components, each corresponding to a step in text-to-SQL dataset annotation and addressing specific user needs mentioned in Section~\ref{sec:user_needs}. 
We describe each component in this section. 
Additional implementation details, such as algorithms and prompt design, are discussed in Appendix~\ref{app:pcfg} and Appendix~\ref{app:prompt}.

\subsection{Schema Visualization}


\begin{figure*}[ht]
  \centering
  \includegraphics[width=\textwidth]{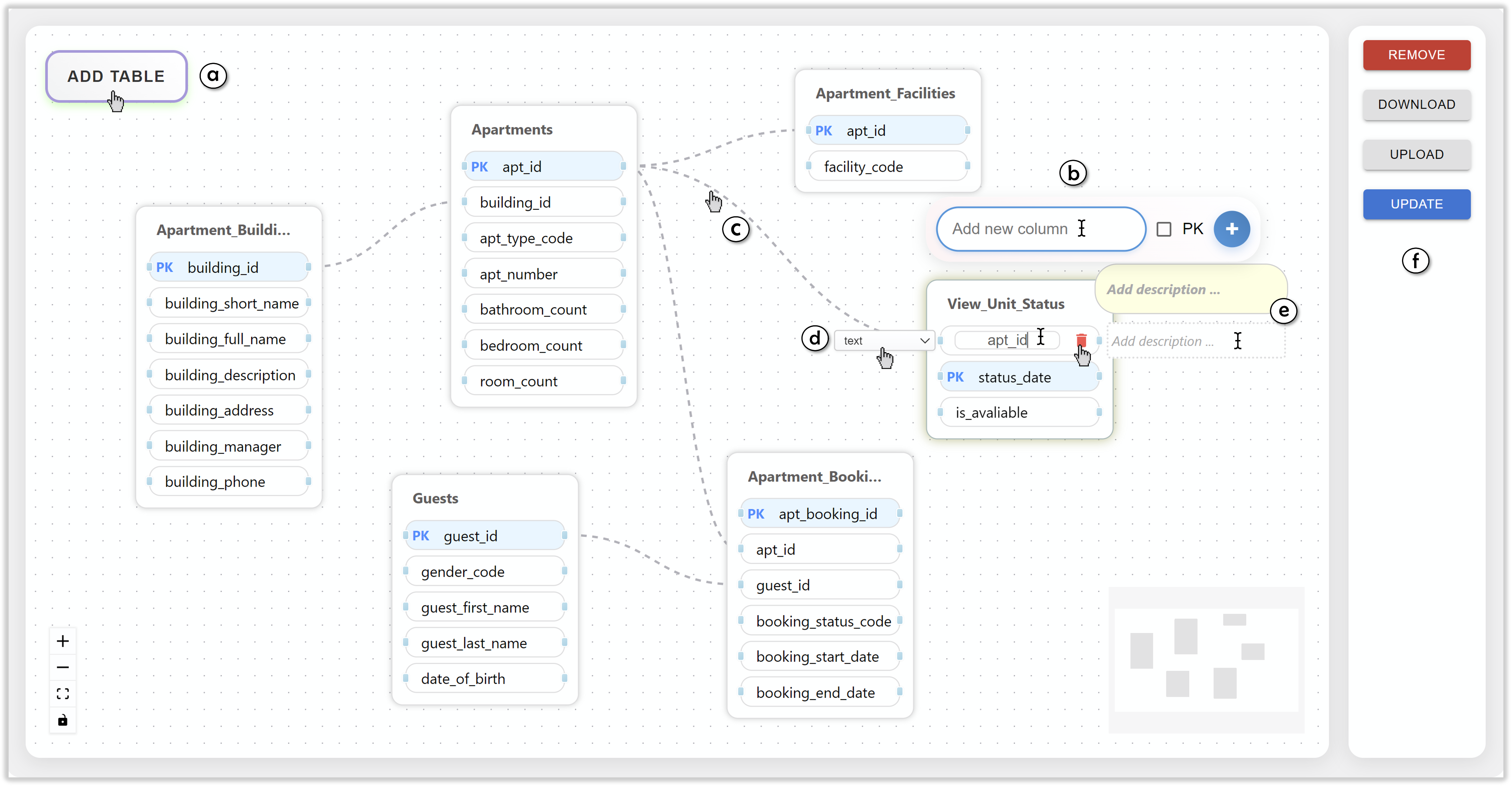}
  \caption{The user interface for schema visualization. \edit{Each node represents a database table, while each cell represents a column in the table. The blue cell marked with ``PK'' represents the primary key. The dashed gray edge represents a foreign key reference relationship between two tables. Users can (a) add a new table, (b) add a new column, (c) add a reference relationship, (d) define the data type for a column, (e) add a description for columns and tables, and (f) remove, upload, or download the database schema.}}
  \label{fig:page1}
\end{figure*}

Annotating text-to-SQL datasets requires annotators to understand the corresponding database schema. However, practical database schemas can be complex and challenging to comprehend. To facilitate user comprehension of schema, {\tool} enables users to visualize the schema in an interactive graph, as shown in Figure~\ref{fig:page1}.
Users can upload the schema by either dragging a schema file onto the canvas or using the upload button. Each table is visualized as a box, with columns listed as rows within it.
The primary keys are colored blue and marked with ``PK''. 
Reference relationships between columns in different tables are rendered as dashed lines, with a flow animation indicating the reference direction.
To inspect details, users can hover over columns and tables to view data types or entity descriptions. The interface allows for dragging tables, zooming in and out, and panning across the view.
 
The graph is editable, allowing users to update the schema as needed. To add new tables, users can click the ``ADD TABLE'' button in the top-left corner (Fig.~\ref{fig:page1} \circled{a}). Hovering over a table allows users to add new columns or designate primary keys (Fig.~\ref{fig:page1} \circled{b}).
Users can specify the data type for each column and directly link two columns to establish foreign key relationships (Fig.~\ref{fig:page1} \circled{c}). To remove columns, users can click the trash icon that appears on hover. 
Tables and reference links can be removed using the backspace key on the keyboard.

Given that practical database entities often use abbreviations, clear documentation can help LLMs better interpret the schema and provide more accurate annotation suggestions in subsequent steps. {\tool} encourages users to add descriptions to tables and columns (Fig.~\ref{fig:page1} \circled{e}).
For better schema management, users can quickly remove the entire schema, as well as download or upload it as a JSON file (Fig.~\ref{fig:page1} \circled{f}).




\begin{figure*}[ht]
  \centering
  \includegraphics[width=0.85\textwidth]{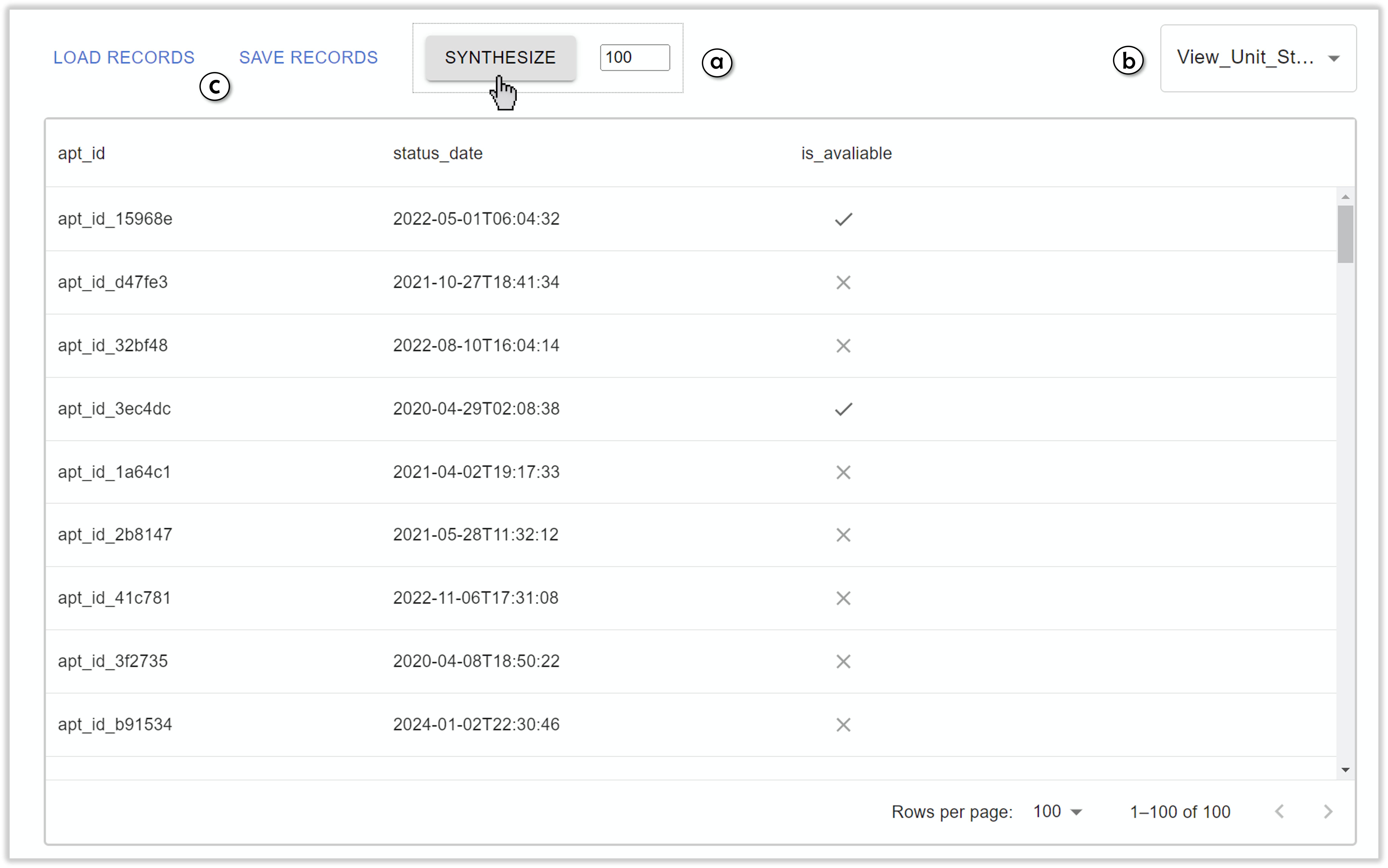}
  \caption{The user interface for database population. \edit{Users can (a) populate the database with a specified number of records, (b) switch table views, and (c) upload or download synthesized records.}}
  \label{fig:page2}
\end{figure*}

\subsection{Database Population}

SQL queries often reference specific values in the database. 
However, there are often no existing records in the database to reference during annotation. 
To address this limitation, {\tool} enables users to instantly create a sandbox database populated with diverse values (Fig.~\ref{fig:page2}). This sandbox database serves two purposes: (1) it provides a source for retrieving values, and (2) it allows for executing the annotated SQL queries to validate their correctness.

To populate the database, users can specify the desired number of records and create them with a single click (Fig.~\ref{fig:page2} \circled{a}). 
{\tool} employs a rule-based method to randomly synthesize records based on data types, which currently supports eight data types, including \texttt{text}, \texttt{boolean}, \texttt{int}, \texttt{timestamp}, \texttt{float}, \texttt{double}, \texttt{decimal}, and \texttt{enum}.
We design random generation rules for each data type. 
For example, in Figure~\ref{fig:page2}, ``apt\_id'' is a text field, and {\tool} generates values by taking ``\textit{apt\_id}'' as the prefix and append a random UUID to it;
``\textit{status\_date}'' is a timestamp field, generating values like ``\textit{2022-05-01T06:04:32}'';
``\textit{is\_available}'' is a boolean field, so it only yields either ``\textit{True}'' or ``\textit{False}'' in the records.

When generating random values, we also consider primary and foreign key relationships to be the constraints.
For example, ``\textit{View\_Unit\_Status.apt\_id}'' is a foreign key referencing another column, ``\textit{Apartments.apt\_id}'', so {\tool} reuses existing values from the referenced column.
The probability of generating repetitive records can be easily adjusted by users in a configuration file, with a default probability of $0.3$.
Users can navigate between different tables via a drop-down menu (Fig.~\ref{fig:page2} \circled{b}). 
Users can save current database records or upload existing ones in JSON format (Fig.~\ref{fig:page2} \circled{c}).


While these rule-based synthetic values may not reflect real-world distributions, this does not affect the annotation process. 
These values primarily act as placeholders for annotators to ensure that the annotated SQL queries can be executed with the desired behavior. 
Furthermore, values are used for reference purposes and do not alter the query's underlying structure or meaning. Models are not directly trained on these values. 


{\tool} distinguishes itself from existing methods that only incorporate dummy values in SQL queries without providing an executable environment.
All SQL queries created by {\tool} are associated with their execution results for users to better understand the query behaviors.




\begin{figure*}[ht]
  \centering
  \includegraphics[width=\textwidth]{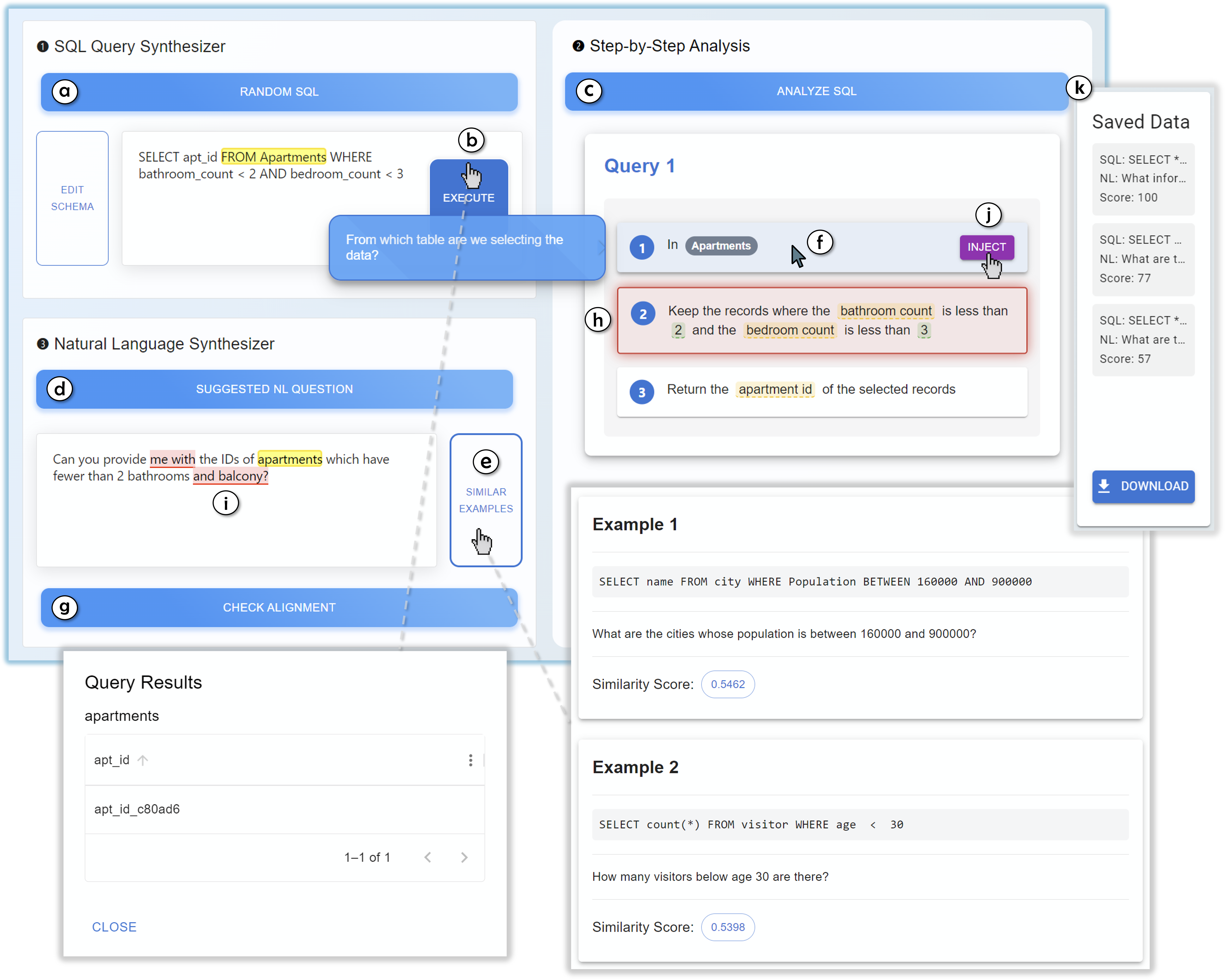}
  \caption{The user interface for data generation, error detection, and repair. \edit{Users can (a) generate a suggested SQL query, (b) check the query result, (c) read the step-by-step explanation in natural language, (d) generate the corresponding suggested NL question, (e) check similar gold data, (f) hover on each step to highlight the corresponding SQL component, NL question chunk, and sub-question, (g) build alignments among SQL, question, and steps, (i) identify and remove redundant text in the question, (j) update the question by emphasizing a certain step, (h) identify a misaligned step, and (k) collect annotated data.}}
  \label{fig:page3_A}
\end{figure*}

\subsection{\textbf{SQL Query Generation}}
\label{sec:sql_sampling}
Creating unbiased SQL queries is challenging, particularly when dealing with a new and complex database schema. 
To address this challenge, {\tool} provides a suggested SQL query (Fig.~\ref{fig:page3_A} \circled{a}) that is randomly sampled using SQL grammar and values in the populated database.
Specifically, {\tool} utilizes a pre-defined probabilistic context-free grammar (PCFG) tailored for SQL queries. This PCFG can be easily modified in a configuration file, as exemplified in Appendix~\ref{app:pcfg}.
While users can configure the grammar manually, {\tool} also supports automatically learning keyword probability distributions from an imported dataset.
Users can directly edit the suggested SQL query to meet specific needs and check the execution result via the ``EXECUTE'' button (Fig.~\ref{fig:page3_A} \circled{b}). 

Compared to using LLMs to generate SQL queries directly, our PCFG-based approach offers more fine-grained control over query diversity and correctness. It mitigates issues such as bias or hallucination introduced by LLMs. This is the rationale behind our decision to first generate the SQL query and then translate it into natural language (NL).
An alternative approach could be generating the NL question first and then generating the SQL query. However, this method has limitations compared to ours. 
First, generating a large amount of diverse NL questions from scratch is challenging. The lack of fixed syntax in NL diminishes the control over data diversity. 
\edit{In contrast, our approach ensures data diversity by directly controlling SQL patterns.}
Second, generating the SQL query from NL essentially implies solving the text-to-SQL task. In this scenario, we can only generate SQL queries by models, which may hallucinate and introduce generation errors in SQL queries~\cite{ning_empirical_2023, tiis_sql}.
\edit{In contrast, the SQL queries generated by our approach are guaranteed to be syntactically correct.}

\subsection{\textbf{Natural Language Question Generation}}
Based on the SQL query, {\tool} provides a suggested NL question by translating the SQL query using GPT-4 Turbo\footnote{All the LLMs mentioned in this paper refer to GPT-4 Turbo (\url{https://platform.openai.com/docs/models/gpt-4-turbo-and-gpt-4}).} (Fig.~\ref{fig:page3_A} \circled{d}).
For more accurate translation, {\tool} employs retrieval-augmented generation  (RAG)~\cite{rag1, rag2}. 
It retrieves similar examples from a text-to-SQL data pool, which collects previously annotated data and 1,500 real-world text-to-SQL pairs. 
\edit{Unlike commonly used retrievers in RAG, such as dense retriever~\cite{dense_retrieve} and BM25~\cite{bm25}, we develop an AST-based retriever tailored for SQL queries.}
\edit{Specifically, {\tool} calculates similarity scores between SQL queries by measuring the tree edit distance between their abstract syntax trees, retrieving the top five examples with scores above $0.5$.}  
Using these examples, {\tool} performs few-shot learning to translate the annotation SQL query into an NL question.
\edit{Figure~\ref{fig:prompt_question} shows the prompt.}
Furthermore, users can view the top similar examples by clicking the ``SIMILAR EXAMPLES'' button (Fig.~\ref{fig:page3_A} \circled{e}). These real-world examples also help users better assess the quality of the LLM-suggested NL question.
The NL question is editable, allowing users to make any necessary adjustments.

\subsection{Error Detection \& Repair}

\subsubsection{SQL Step-by-step Explanation in NL}
To enhance user comprehension of SQL queries and detect potential errors, {\tool} explains the SQL query step by step in NL (Fig.~\ref{fig:page3_A} \circled{c}).
We reuse the rule-based explanation generation approach from STEPS~\cite{STEPS}, which parses the SQL query and translates each part of the query to an NL description based on templates.
{\tool} enhances this approach using an LLM in two ways. 
First, if a SQL query cannot be fully covered by the translation rules, {\tool} prompts the LLM to generate the step-by-step explanation via few-shot learning from rule-based explanation examples. 
Second, {\tool} prompts the LLM to paraphrase the generated explanations based on the schema and documentation for better fluency and naturalness. \edit{Figure~\ref{fig:prompt_explanation} shows the prompt.} To further improve the readability, {\tool} identifies and highlights columns, tables, and values in different colors in the explanation. 
Furthermore, for each step, {\tool} renders a corresponding sub-question on the left as users hover the mouse over this step (Fig.~\ref{fig:page3_A} \circled{f}). 
The sub-question is translated from a simple SQL query that only involves this step.

\subsubsection{Visual Correspondence among SQL query, NL question, and Step-by-step Explanation}
The step-by-step explanation serves as a bridge between the SQL query and the NL question. When users click the ``CHECK ALIGNMENT'' button (Fig.~\ref{fig:page3_A} \circled{g}), {\tool} creates a triple-linkage among these elements.
First, since the step-by-step explanation is grammar-based, there is a one-to-one mapping between SQL components and explanation steps.
Second, {\tool} employs an LLM to align the step-by-step explanation with the NL question. For each explanation step, the LLM pinpoints related substrings in the NL question, maintaining a one-to-many mapping.
When users hover over an explanation step (Fig.~\ref{fig:page3_A} \circled{f}), {\tool} highlights the corresponding SQL component and related question substrings in yellow.
This triple-linkage helps users mentally connect the SQL query and the suggested NL question, enhancing user understanding of the data and aiding in the detection of potential errors.

\subsubsection{Misalignment Detection \& Correction}

While {\tool} guarantees the syntactic correctness of SQL queries sampled by PCFG, the NL question generated by the LLM can include errors or ambiguity.
{\tool} proposes a novel interactive error detection and correction strategy by aligning the NL question with the SQL query through the step-by-step explanation.
\edit{Motivated by research~\cite{multi_agent_collaboration} showing that multi-agent collaboration enhances generation accuracy, {\tool} accomplishes this task through a two-step prompting.}
\edit{We include our prompt design in Figures~\ref{fig:prompt_alignment_analysis} and~\ref{fig:prompt_alignment_map}, with further details discussed in Appendix~\ref{app:prompt}.}
If any substring in the NL question fails to map to any step in the explanation, the substring will be highlighted in red (Fig.~\ref{fig:page3_A} \circled{i}), suggesting that this text may be irrelevant to this SQL query.
Users can focus on the red text and consider removing them.
Similarly, if a certain explanation step does not map to any partial text in the NL question, this step will be highlighted in red (Fig.~\ref{fig:page3_A} \circled{h}), indicating the content mentioned in this step may be missing in the NL question.
In this case, users can update the NL question by clicking the ``INJECT'' button on the corresponding step (Fig.~\ref{fig:page3_A} \circled{j}). Then, the LLM is prompted to update the current NL question by amplifying the content mentioned in this step.
\edit{Figure~\ref{fig:prompt_inject} shows the prompt.}

\subsubsection{Confidence Scoring}
\begin{figure*}[ht]
  \centering
  \includegraphics[width=\textwidth]{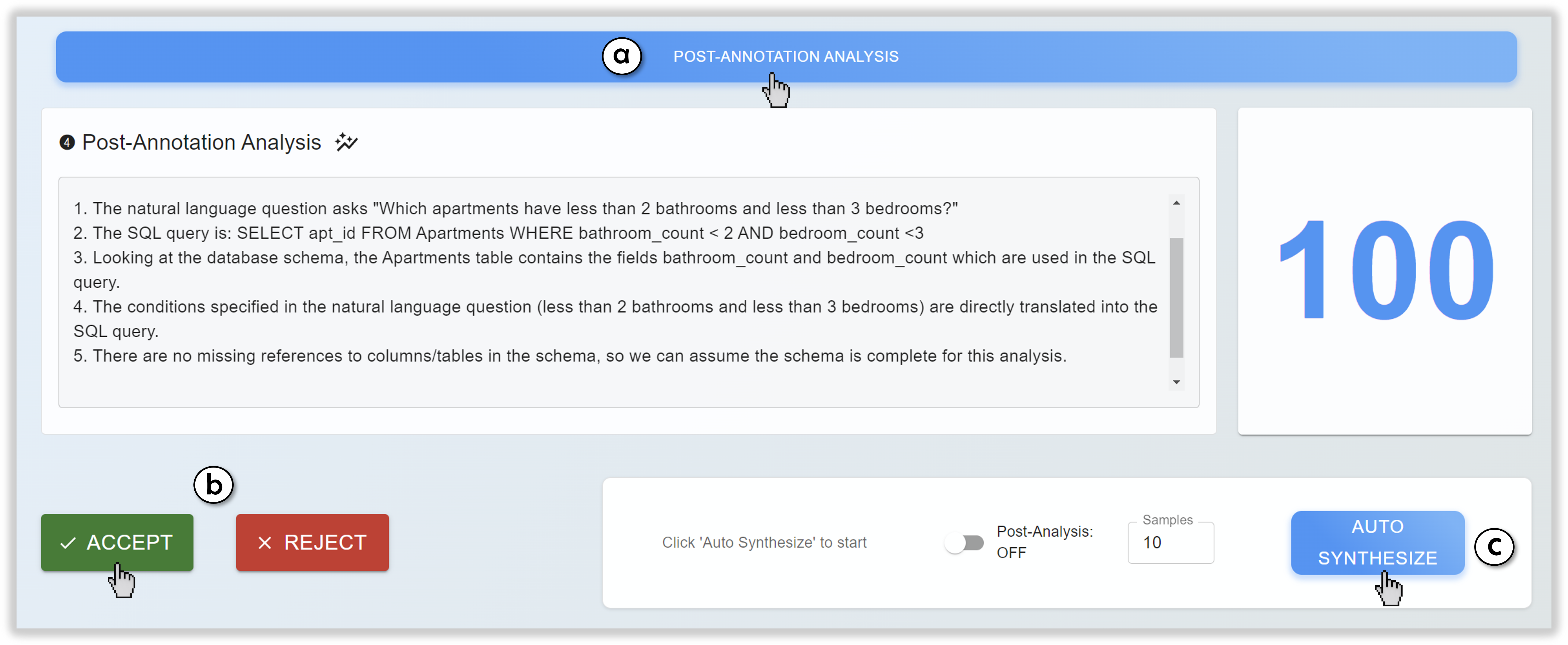}
  \caption{The user interface for post-synthesis analysis \& automated annotation. \edit{Users can (a) generate an analysis report and scoring for annotating the current data pair, (b) accept or reject the current data pair, and (c) start automated data annotation without human intervention.}}
  \label{fig:page3_B}
\end{figure*}

To help users better assess the quality of annotated data, {\tool} offers a post-annotation analysis via the ``POST-ANNOTATION ANALYSIS'' button (Fig.~\ref{fig:page3_B} \circled{a}). Recent research has demonstrated that LLMs can determine semantic equivalence between SQL queries~\cite{llm_sql_equivalence} and generate accurate confidence scores through self-reflection~\cite{calibration_and_correctness, tian2023justaskcalibrationstrategies}.
Based on these findings, {\tool} prompts the LLM to provide a final report and score indicating the quality and correctness of the data.  \edit{The prompt used in this step is shown in Figure~\ref{fig:prompt_equivalence}.}
The score is averaged after multiple rounds of analysis to ensure stability. 
This score serves as a confidence level, directing users to focus more on checking data with lower scores, as these data may contain potential issues.

Based on the analysis provided by {\tool}, users can choose to accept or reject the data (Fig.~\ref{fig:page3_B} \circled{b}). Accepted data is collected in the right panel (Fig.~\ref{fig:page3_A} \circled{k}), where users can review and download the dataset at any time.

\subsection{Automated Dataset Annotation}
While {\tool} enables users to annotate text-to-SQL datasets in an interactive manner, {\tool} also supports fully automated data annotation without humans in the loop (Fig.~\ref{fig:page3_B} \circled{c}). This is useful when users need a large amount of data without a perfect dataset quality (e.g., for fine-tuning LLMs).
Users can specify how many queries to synthesize and start by one click.
All generated data will be automatically collected on the right (Fig.~\ref{fig:page3_A} \circled{k}).

\subsection{Dataset Diversity Analysis}

\begin{figure*}[ht]
  \centering
  \includegraphics[width=\textwidth]{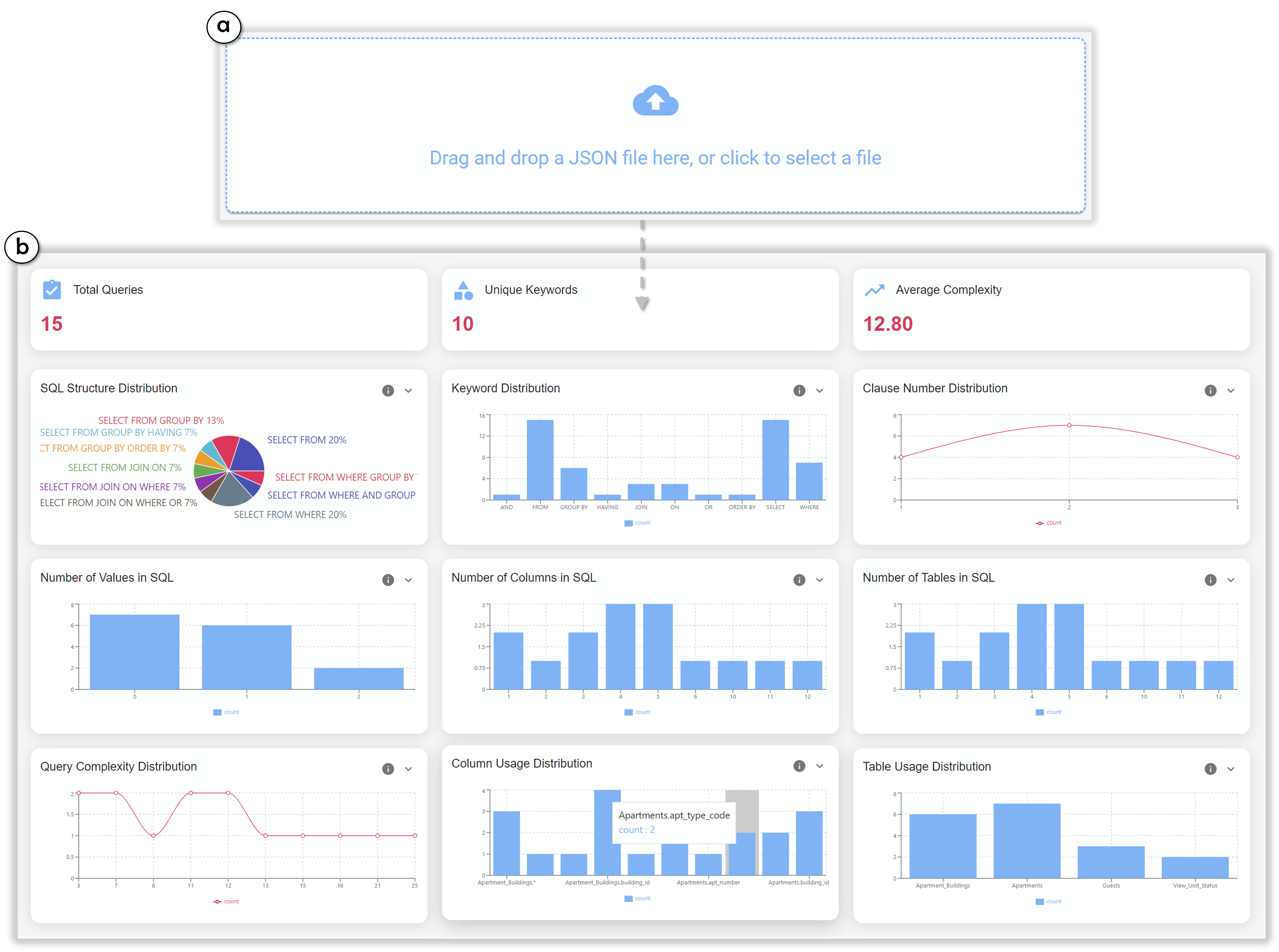}
  \caption{The user interface for dataset diversity analysis. \edit{Users can (a) upload existing annotated text-to-SQL datasets and (b) monitor various property distributions.}}
  \label{fig:page4}
\end{figure*}

To ensure diversity and eliminate potential biases in the annotated dataset, {\tool} allows users to monitor dataset composition and property distributions.
Users can upload their dataset via drag-and-drop (Figure~\ref{fig:page4} \circled{a}). {\tool} then renders various property distributions in pie charts, bar charts, or line charts, in terms of SQL structure, keyword, clause number, column usage, etc. (Figure~\ref{fig:page4} \circled{b}).
For example, users can monitor the number of referenced values in a bar chart.
If users find that SQL queries with a sufficient number of referenced values are underrepresented in the current dataset, they can adjust the PCFG probabilities to generate SQL queries with more values. And they can selectively accept only those queries that contain an adequate number of values.
In addition to ensuring diversity, this UI component generally improves human control during collaboration with the LLM, enabling users to better manage annotation pace and focus.

\section{Usage Scenario}

Bob is a data scientist at a rapidly growing technology company. 
Now, his task is to create a high-quality text-to-SQL dataset for training and evaluating the natural language (NL) interface of the company's recently updated database system.
Bob faces several challenges that make this task particularly daunting.
First, the company has just completed a major update to its database schema, introducing new tables and relationships to accommodate its expanding business needs. 
This update makes previous datasets obsolete and incompatible with the current schema. 
Consequently, it is impossible to accurately evaluate the performance of the NL interface based on the updated database.
Adding to the complexity, the schema now becomes highly complex, with numerous tables and reference relationships. Manually updating previous datasets to reflect these changes would be impractical. Bob realizes that he needs a solution that can handle this complexity efficiently and accurately.
Furthermore, Bob needs to create diverse, unbiased SQL queries and their corresponding NL questions at scale. Doing this manually would be prohibitively time-consuming and challenging, especially given the complexity of the new schema.
Recognizing these challenges, Bob decides to use the newly developed text-to-SQL data annotation tool, {\tool}, to streamline his workflow and ensure the creation of a controllable, high-quality dataset.

\textbf{Schema Comprehension.}
Bob begins by uploading a JSON file containing the company's updated database schema to {\tool}. As the schema loads onto the drag-and-drop canvas, Bob is immediately impressed by how {\tool} transforms the complex JSON structure into an intuitive visual representation. Tables appear as clearly defined boxes with columns listed inside, while relationships between tables are displayed as animated dashed lines.
The visual layout allows Bob to quickly grasp the overall structure of the database, saving him hours of time that would have been spent mentally parsing the JSON file.
Using the intuitive interface, Bob makes necessary adjustments. He double-clicks to edit table names, drags lines to establish reference relationships, and documents the meaning of an abbreviated column name. The ability to zoom in and out further helps Bob navigate the complex structure. 
As he works, Bob realizes the significant improvement in efficiency compared to editing the schema through the original schema definition file directly. What might have taken hours of painstaking work is now being accomplished in minutes, with much greater accuracy and confidence. Finally, Bob downloads the updated schema. He feels confident that this new well-documented schema will serve as a valuable foundation for future projects.

\textbf{Database Population.}
For more convenient data annotation, Bob populates the database with synthetic records with {\tool}. He specifies a need for 1,000 employee records. Upon clicking the SYNTHESIZE button, {\tool} instantly creates these records. Bob reviews the generated data and notices that the synthetic employee names look diverse. He proceeds to generate records for the other tables. Satisfied with the data generation, Bob downloads the database for future use and moves on to the next step.

\textbf{Data Annotation.}
Given the database, Bob is ready to annotate text-to-SQL data by creating SQL queries and their corresponding NL questions.
Bob finds manually creating a SQL query from scratch challenging, so he decides to generate a random SQL query by {\tool}:

\begin{center}
\texttt{\textcolor[RGB]{172,41,0}{SELECT} Employees.name} \texttt{\textcolor[RGB]{172,41,0}{FROM} Employees} \\
\texttt{\textcolor[RGB]{172,41,0}{WHERE} Employees.department\_id = 5 \textcolor[RGB]{172,41,0}{AND} Employees.salary > 50000}
\end{center}

Bob finds the SQL query reasonable. To confirm his understanding, Bob clicks the ANALYZE SQL button, and {\tool} shows a step-by-step analysis:

\begin{enumerate}
    \item \texttt{\textcolor[RGB]{172,41,0}{FROM} Employees} $\rightarrow$ \textit{Which data source should we care about?} 
          \\ \colorbox[rgb]{0.95,0.95,0.95}{In employees}

    \item \texttt{\textcolor[RGB]{172,41,0}{WHERE} Employees.department\_id = 5} $\rightarrow$ \textit{Which department are employees from?} 
          \\ \colorbox[rgb]{0.95,0.95,0.95}{Filter employees from department 5}

    \item \texttt{\textcolor[RGB]{172,41,0}{AND} Employees.salary > 50000} $\rightarrow$ \textit{What salary range do we care about?} 
          \\ \colorbox[rgb]{0.95,0.95,0.95}{Keep employees with salary exceeding \$50,000}

    \item \texttt{\textcolor[RGB]{172,41,0}{SELECT} Employees.name} $\rightarrow$ \textit{What information should be returned?} 
          \\ \colorbox[rgb]{0.95,0.95,0.95}{Return the names of employees} 
        
\end{enumerate}

As Bob hovers over each step, a corresponding sub-question is rendered in the tooltip and the corresponding SQL component is highlighted.
{\tool} then generates a suggested NL question for this query:
\begin{center}
``\textit{\textbf{Who are the employees in the marketing department with a salary higher than \$50,000 and have been with the company for over 5 years?}}''
\end{center}
However, Bob notices that this question does not perfectly match the SQL query and decides to use the alignment feature to refine it.
Bob clicks the CHECK ALIGNMENT button, eager to see how well the generated question matches the SQL query.
He is immediately drawn to a phrase in the question highlighted in red: ``\textit{marketing department}'', suggesting there is no corresponding element in the SQL query.
Bob realizes this information is irrelevant and needs to be removed.
To better understand the quality of this suggested query, Bob hovers over the step-by-step explanation. To his surprise, {\tool} further visually corresponds each explanation step to sub-strings of the NL question through simultaneous highlighting.
He notices one explanation step, ``\textit{Filter employees from department 5}'', is highlighted in red.
This visual cue tells Bob that this step is not reflected in the current question.

Bob decides to address these issues one by one. 
First, he removes the irrelevant information by deleting the red-highlighted phrase ``\textit{marketing}'' and the unrelated condition ``\textit{and have been with the company for over 5 years}'' from the question.
Next, he turns his attention to the missing information about the department. He hovers over the red-highlighted explanation step, ``{\em Filter employees from department 5}'', and an INJECT button appears. Bob clicks this button and the current NL question is updated by incorporating this step.
The question now becomes:
\begin{center}
\textit{\textbf{Who are the employees in Department 5 with a salary higher than \$50,000?}}
\end{center}
Excited to see the results of his edits, Bob clicks the CHECK ALIGNMENT button again. 
This time, Bob notices that there is no red highlight in either the explanation or the NL question.
As a final check, Bob hovers over the explanation steps. He watches with satisfaction as each step successfully maps to a SQL component and sub-strings in the NL question.

\begin{table*}[htbp]
\centering
\small
\begin{tabularx}{\textwidth}{>{\raggedright\arraybackslash}p{0.33\textwidth}>{\raggedright\arraybackslash}p{0.33\textwidth}>{\raggedright\arraybackslash}X}
\toprule
\textbf{Explanation Step} & \textbf{SQL Query Component} & \textbf{Question Sub-string} \\
\midrule
(1) In employees & \texttt{\textcolor{brown}{FROM} Employees} & \textit{... the employees ...} \\
(2) Filter employees from department 5 & \texttt{\textcolor{brown}{WHERE} Employees.department\_id = 5} & \textit{... in department 5 ...} \\
(3) Keep employees with salary exceeding \$50,000 & \texttt{\textcolor{brown}{AND} Employees.salary > 50000} & \textit{... with a salary higher than \$50,000...} \\
(4) Return the names of employees & \texttt{\textcolor{brown}{SELECT} Employees.name} & \textit{Who are the employees ...} \\
\bottomrule
\end{tabularx}
\label{tab:mapping}
\end{table*}

According to the visual alignment, Bob is confident that the NL question matches the SQL query. He further validates it by clicking the POST-SYNTHESIS button. 
{\tool} then reports the equivalence analysis in a short paragraph, along with a high confidence score of 98. Pleased with the result, Bob accepts this text-to-SQL instance and collects it to the right panel. He appreciates how the interactive alignment feature and intuitive triple-linkage visualization help him efficiently identify and correct misalignments with high confidence in the data's quality.

As Bob progresses, he periodically uses {\tool} to analyze the dataset composition to ensure he creates a diverse and balanced dataset. He notices that queries involving the newly added tables and relationships are underrepresented, so he adjusts the query generation parameters to increase their frequency.
By the end of the day, Bob creates a substantial, high-quality text-to-SQL dataset that accurately reflects the company's updated database schema. This new dataset will be invaluable for both training and evaluating their natural language interface, something that was not possible before due to the lack of relevant evaluation data.
Bob feels a sense of accomplishment. He successfully updates and documents a complex schema that would have been extremely time-consuming and error-prone to modify manually. He creates a dataset specific to the company's current database schema, including new tables, columns, and relationships. More importantly, the dataset provides a strong evaluation benchmark for the updated schema, allowing the team to accurately evaluate the performance of their NL interface.
The interactive nature of the tool allows Bob to leverage his domain knowledge while benefiting from automated generation and analysis features. He appreciates how the tool transforms a typically tedious and challenging process into an efficient and engaging one, ultimately contributing to the improvement of the company's data interaction capabilities.
\section{User Study}
To investigate the usability and effectiveness of {\tool}, we conducted a within-subjects user study with 12 participants. The study compared {\tool} with manual annotation and the use of a conversational AI assistant.

\subsection{Participants}
We recruited 12 participants (4 female, 8 male) from Adobe. They worked in different roles including Machine Learning Engineers, Research Scientists, Data Scientists, and Product Managers.
Their works were directly or indirectly related to querying data in the database.
All of them had either Master's or PhD degrees.
Participants self-rated their proficiency in SQL (\textit{3 Beginner}, \textit{3 Basic}, 4 \textit{Intermediate}, \textit{2 Advanced}) and usage frequency of LLMs (6 \textit{Yearly}, 2 \textit{Monthly}, 2 \textit{Weekly}, 3 \textit{Daily}).

\subsection{Tasks}

We randomly sampled 9 schemas on the widely used text-to-SQL benchmark, Spider~\cite{spider}. 
We provided these schemas in JSON format, whose syntax was comprehensible to all participants.
Based on the schema, participants were asked to annotate text-to-SQL data while optimizing both the data quantity and quality.


\subsection{Comparison Baselines}
To the best of our knowledge, no text-to-SQL data annotation tools were readily available for comparison at the time of the user study. Thus, we compared {\tool} with two commonly applied scenarios for text-to-SQL dataset annotation in the industry, manual annotating and using an AI assistant. 

\noindent \textbf{Manual.} We asked participants to manually review and customize the schema, create SQL queries, and write corresponding NL questions. They recorded the results in an Excel sheet. 

\noindent \textbf{AI Assistant.} We gave participants access to using a state-of-the-art conversational AI assistant, ChatGPT (GPT-4). 
For example, participants could directly upload the entire schema file in their conversation with ChatGPT and request the generation of sufficient text-to-SQL data.
We did not impose any restrictions on how participants should use ChatGPT.

\subsection{Protocol}
Each study began with a demographic survey and study introduction. Participants then watched a 4-minute tutorial video of {\tool} and spent 3 minutes practicing to get familiar with it. Meanwhile, we collected quality feedback from users.

After participants were familiar with {\tool}, they proceeded to annotate in the assigned condition (i.e., Manual, AI assistant, {\tool}). Each task consisted of three 5-minute sessions, one for each condition. 
We randomized the order of assigned conditions to mitigate learning effects.
For each session, participants were provided with a database schema and asked to annotate as many text-to-SQL datasets as possible. We asked participants to focus on not only the quantity but also the quality of annotated data.

After each session, participants completed a post-task survey to rate their experience with the assigned condition. The survey included the System Usability Score (SUS)~\cite{sus} and NASA Task Load Index (TLX)~\cite{NASA-TLX} questionnaires, using 7-point Likert scales to assess their perceptions. At the end of the study, participants filled out a final survey sharing their experiences, opinions, and thoughts. The entire study took approximately 70 minutes.

\section{Results}

This section describes the results of our user study. 
We first present the analysis of user annotation performance in different conditions. We measure annotation speed and annotation quality.
Then, we report user perception of different conditions, e.g., their confidence level of annotated data and the perceived cognitive load. 

\subsection{Annotation Speed}

\begin{table}[htb]
    \centering
    \caption{Number of Annotated Data (5 minutes)}
    \vspace{-2.5mm}
    \resizebox{0.8\linewidth}{!}{%
    \begin{tabular}{lcc}
        \toprule
              & \textbf{Completed Annotation} & \textbf{SD} \\
        \midrule
        Manual & 2.00 & 0.91 \\
        AI Assistant & 3.75 & 2.09 \\
        {\tool} & \textbf{8.75} & 2.74 \\
        \bottomrule
    \end{tabular}
    }
    \label{tab:annotation_speed}
\end{table}

Since each session has a fixed annotation time period, we use the average number of annotated tasks to represent the annotation speed. 
We compare the number of tasks completed across three conditions: Manual, AI assistant, and {\tool}. Table~\ref{tab:annotation_speed} presents the average annotation count completed within 5 minutes of the task session for each condition.
When using {\tool}, participants annotated the most tasks ($Mean = 8.75, SD = 2.74$), followed by using the AI assistant ($Mean = 3.75, SD = 2.09$), and manually annotating ($Mean = 2.00, SD = 0.91$). The ANOVA test shows that the mean differences are statistically significant ($p$-value = 1.96e-8).

The substantial improvement in task achievement with {\tool} suggests that {\tool} could enhance productivity in real-world applications.
Compared to ChatGPT, {\tool} provides a better explanation for participant engagement and validation.
P3 said, ``\textit{Even though ChatGPT can give me plenty of data, I need to manually check its output. I feel that connecting its output SQL to the schema is challenging in the limited time.}''
Furthermore, {\tool} serves as an interactive interface, allowing users to iteratively refine the data.
Both P6 and P9 appreciated this utility.
P9 said, ``\textit{It's easier and quicker to test and iterate on these [data] using the tool}'',

\subsection{Annotation Quality}
We define the quality of a text-to-SQL dataset into (1) \textbf{Correctness} (whether there is a syntax error, and whether the NL question matches the SQL query), (2) \textbf{Naturalness} (whether the NL question is natural enough as a human daily question), and (2) \textbf{Diversity} (whether the dataset has comprehensive coverage of different entities and query types, without any biases).

\subsubsection{\textbf{Correctness}}

To evaluate the correctness of participants' annotations, we manually review all data collected during the user study. We evaluate two types of errors in the annotated data. 
First, we look for SQL syntax errors or the misuse of entities and references in the database schema. 
We identify this type of error by executing the SQL query in a sandbox database that is adequately populated from the schema.
The error leads to execution failures or the return of an empty result.
Second, we evaluate the equivalence between the SQL query and the NL question. While the SQL might be syntactically correct on the schema, it may not accurately represent the intent of the NL question. In this case, we manually evaluate their equivalence.
Table~\ref{tab:correctness_annotation} shows the two types of error rates and the overall accuracy for each condition.

We observe different reasons for errors in manual annotation compared to using the AI assistant. During manual annotation, participants who were less proficient in SQL often made grammatical mistakes (29.24\%). However, since they tended to write simple SQL queries, the equivalence error was less (5.15\%).
The AI assistant, ChatGPT, rarely introduces syntax errors. However, it tends to generate more complex SQL queries (e.g., multiple JOINs) than manual annotation. Despite participant refinement, these queries often fail to match the complex schema due to hallucination, leading to SQL execution errors (18.73\%). Moreover, these complex queries pose greater challenges in maintaining equivalence with the NL question, resulting in the highest error rate of equivalence ((8.34\%).

Compared to manual annotation and using the AI assistant, using {\tool} achieves the highest accuracy (95.56\%). For SQL errors, queries sampled by {\tool} are guaranteed to be syntactically correct. Regarding NL-SQL equivalence, {\tool} aligns SQL and NL through step-by-step analysis, achieving the best equivalence after human refinement. However, we observe cases (4.44\%) where they are not fully equivalent, suggesting room for further improvement in NL generation accuracy.

\begin{table}[htb]
    \centering
    \caption{Correctness of Annotated Data}
    \vspace{-2.5mm}
    \resizebox{1\linewidth}{!}{%
    \begin{tabular}{lccc}
        \toprule
              & \textbf{SQL Error} & \textbf{Equivalence Error} & \textbf{Accuracy} \\
        \midrule
        Manual  & 29.24\%  & 5.15\% & 65.61\% \\
        AI Assistant   & 18.73\%   & 8.34\% & 72.93\% \\
        {\tool} & \textbf{0}   & \textbf{4.44\%} & \textbf{95.56\%} \\
        \bottomrule
    \end{tabular}
    }
    
    \label{tab:correctness_annotation}
\end{table}

\subsubsection{\textbf{Naturalness}}
In addition to correctness, the naturalness of the NL question is crucial for the quality of text-to-SQL data.
While an NL question may accurately match its SQL query, it might be verbose. In the real world, people tend to ask concise questions that follow certain natural language patterns. To evaluate naturalness, we first calculate the \textit{Flesch-Kincaid Readability Score}~\cite{flesch_score}, an automatic metric measuring text readability on a scale from 0 to 100.
To better assess naturalness, we further manually rate all annotated questions from 1 to 7 after masking the conditions.\footnote{{\tool} references synthetic values in a sandbox database, which can be easily identified by the human raters. We replace all these values with commonly used values for fair evaluation.}

\begin{table}[htb]
    \centering
        \centering
        \caption{Flesch-Kincaid Readability Score of Annotated Questions (0-100).}
        \vspace{-2.5mm}
        \resizebox{0.69\linewidth}{!}{%
        \begin{tabular}{lcc}
            \toprule
                  & \textbf{Flesch-Kincaid Score} & \textbf{SD} \\
            \midrule
            Manual & 76.94 & 13.15 \\
            AI Assistant & 56.14 & 18.62 \\
            {\tool} & 72.32 & 16.54 \\
            \bottomrule
        \end{tabular}
        }
        \label{tab:Flesch-Kincaid}
\end{table}

\begin{table}[htb]
    \hspace{0.06\textwidth}
        \centering
        \caption{Manual Rating of Naturalness of Annotated Questions (0-7).}
        \vspace{-2.5mm}
        \resizebox{0.65\linewidth}{!}{%
        \begin{tabular}{lcc}
            \toprule
                  & \textbf{Human Rating} & \textbf{SD} \\
            \midrule
            Manual & 6.25 & 0.52 \\
            AI Assistant & 6.02 & 0.58 \\
            {\tool} & 6.19 & 0.44 \\
            \bottomrule
        \end{tabular}
        }
        \label{tab:natural_rating}
\end{table}

\begin{table*}[htb]
    \centering
    \caption{SQL Query Component Diversity Analysis in Annotated Datasets.}
    \vspace{-2.5mm}
    \resizebox{0.75\linewidth}{!}{%
    \begin{tabular}{l cc|cc|cc|cc}
        \toprule
        & \multicolumn{2}{c}{\textbf{Clause}} & \multicolumn{2}{c}{\textbf{Table}} & \multicolumn{2}{c}{\textbf{Column}} & \multicolumn{2}{c}{\textbf{Value}} \\
        \cmidrule(lr){2-9}
        \textbf{Method} & Diversity & Mean & Diversity & Mean & Diversity & Mean & Diversity & Mean \\
        \midrule
        Manual       & 0.48 & 2.40 & 0.41 & 1.29 & 0.17 & 1.09 & 0.44 & 0.33 \\
        AI Assistant & 0.64 & 5.10 & 0.81 & 5.75 & \textbf{0.64} & 3.73 & 0.62 & 1.25 \\
        {\tool}      & \textbf{0.69} & 3.18 & \textbf{0.83} & 2.54 & 0.59 & 1.71 & \textbf{0.66} & 0.85 \\
        \bottomrule
    \end{tabular}
    }
    \label{tab:diversity}
\end{table*}

\begin{figure*}[htb]
  \centering
\includegraphics[width=0.85\textwidth]{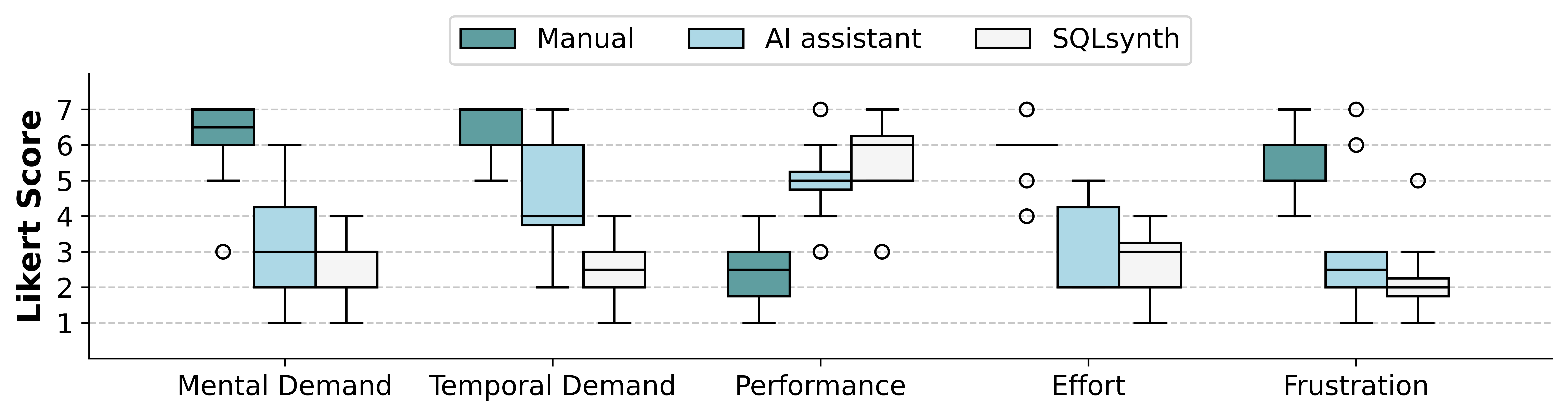}
  \caption{NASA Task Load Index Ratings of Text-to-SQL Data Annotation}
  \label{fig:cognitive1}
\end{figure*}

\begin{figure*}[htb]
  \centering
  \includegraphics[width=0.92\textwidth]{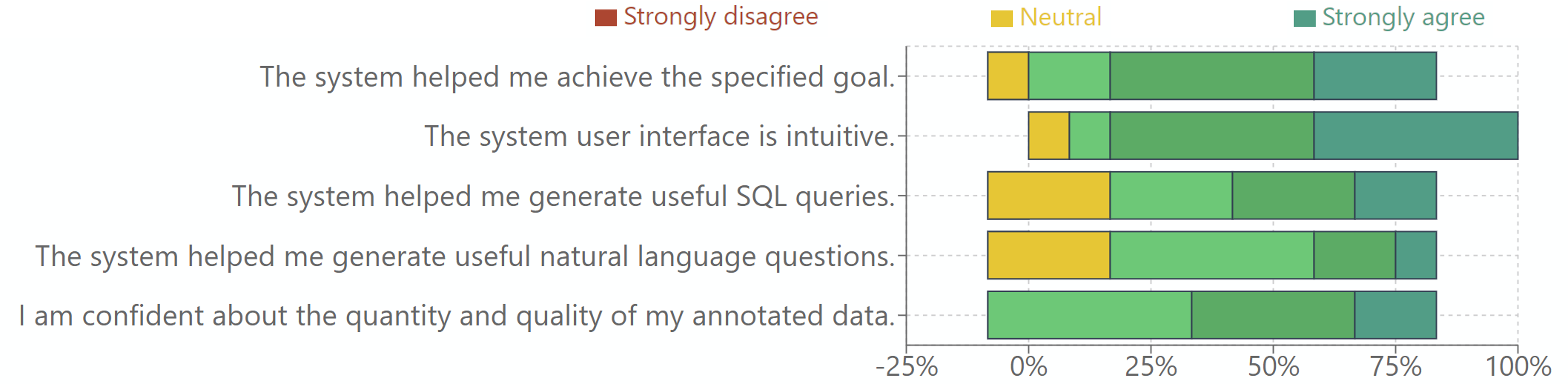}
  \caption{SUS Scores of Text-to-SQL Data Annotation}
  \label{fig:sus1}
\end{figure*}

As shown in Table~\ref{tab:Flesch-Kincaid} and Table~\ref{tab:natural_rating}, the \textit{Flesch-Kincaid Readability Score} is consistent with human ratings. While manual annotation achieves the highest score (Flesch-Kincaid Score: 76.94, Human Rating: 6.25) as expected, the NL questions annotated through {\tool} achieve a comparable score (Flesch-Kincaid Score: 73.32, Human Rating: 6.19). We observe that using ChatGPT achieves the worst naturalness in questions (Flesch-Kincaid Score: 56.14, Human Rating: 6.02). Notably, ChatGPT-generated questions often include SQL keywords or follow SQL query patterns (e.g., \textit{Return all student names that are grouped by grades.}). 
Participants often accept ChatGPT's generation without modifications. While {\tool} is also built upon ChatGPT, it employs step-by-step analysis and in-context learning from real-world questions, thereby better incorporating real question patterns into the generated questions. 
Moreover, {\tool} offers a better interface and provides helpful suggestions for refining questions. Participants report that they are more willing to polish the LLM-generated data when using {\tool}.
P8 said, ``\textit{I really enjoy the UI in this tool. It suggests how to polish the data. And I really enjoy playing with the alignment feature to see what I can do with existing data.}''

\subsubsection{\textbf{Diversity}}
To evaluate the diversity and potential biases in the annotated dataset, we analyze the composition of participant-annotated data across four dimensions, including the number of clauses, columns, tables, and values involved. We measure the diversity of each dimension using Simpson's Diversity Index~\cite{simpson_diversity}, which is used to quantify the level of heterogeneity of a certain property.

Table~\ref{tab:diversity} shows the diversity and mean values for each method and dataset property. {\tool} demonstrates better diversity in the generated SQL for most dimensions, except for columns generated by the AI assistant. 
This is because ChatGPT tended to join excessive tables (mean = 5.75) in \texttt{FROM} clauses and include excessive columns (mean = 3.73) in  \texttt{SELECT} clauses.
In contrast, {\tool} learned the distribution from real-world datasets, resulting in a more reasonable distribution. For manual annotation, participants typically wrote simple SQL queries. For instance, they rarely used JOIN clauses, leading to low diversity.

\subsection{User Cognitive Load \& Usability Rating}


Reducing cognitive load during data annotation is crucial, since it directly makes this process more cost-effective. Figure~\ref{fig:cognitive1} illustrates participants' ratings on the five cognitive load factors from the NASA TLX questionnaire. The ANOVA test reveals statistically significant ($\alpha = 0.001$) differences in means for all factors: Mental Demand ($p = 6.7 \times 10^{-5}$), Temporal Demand ($p = 6.0 \times 10^{-6}$), Performance ($p = 1.9 \times 10^{-5}$), Effort ($p = 1.5 \times 10^{-5}$), and Frustration ($p = 1.3 \times 10^{-4}$).

The result demonstrates that {\tool} can significantly reduce users' cognitive load compared to manual annotation and using ChatGPT.
P1 said, ``\textit{Generating NL2SQL data felt effortless with the help of the tool.}''.
We believe this is achieved by a smoother collaboration between the huamn and the LLM.
P1 also comprehensively discussed how {\tool} reduces users' cognitive load,
``\textit{The system allows me to generate a variety of SQLs without any cognitive efforts. This was great because I didn't have to think about SQL syntax, different queries one might ask about the dataset. I also like that the system can generate corresponding NL questions for each SQL. While the generated NL question wasn't always accurate, the system already provided me something which I could iterate on. This is almost like someone gives you a draft that you can just revise vs.~gives you an empty doc for you to start from scratch in writing.}''


Figure~\ref{fig:sus1} displays SUS scores reported by participants, showing consistently positive feedback across all dimensions. Notably, no participants disagreed with any dimension. While a few participants expressed neutral opinions about SQL queries and NL questions suggested by {\tool}, most viewed them positively. All participants expressed high confidence in the quality and quantity of data annotated using {\tool}.
\section{Discussion}

\subsection{Design Implications}

{\tool} demonstrates significant performance improvements compared to manual annotation and using ChatGPT. Based on participants' feedback, we attribute the improvement to the four key design ideas of {\tool}: (1) Structured workflow, (2) Data suggestion, (3) Error detection and repair, and (4) Data visualization. 

First, manual annotation and AI-assisted approaches lack structure and standards. In contrast, {\tool} structures the workflow into concrete steps and provides clear guidance (e.g., the next component blinks in the UI).
P11 wrote, ``\textit{Overall, the tool is extremely useful and facilitates the annotation, which is an admittedly open task. It adds structure to an unstructured task.}''
Second, the grammar-based suggested SQL queries and LLM-based NL questions serve as unbiased, reliable starting points for annotation. It enables annotators to avoid the most challenging and biased step. Instead of creating data from scratch, annotators only need to refine the suggested ones.
P3 wrote, ``\textit{Even though ChatGPT can give me plenty of data, I need to manually check its output. I feel connecting its output SQL to the schema is challenging given the limited time.}''
P5 wrote, ``\textit{It is useful for generating synthetic question and SQL pairs.}''
P1 wrote, ``\textit{This is almost like someone gives you a draft that you
can just revise vs. gives you an empty doc for you to start from scratch in writing.}''
Third, {\tool} highlights potential errors according to the misalignment between SQL and NL, saving effort on error checking and correction. 
P6 wrote, ``\textit{I could see how the alignment feature could be useful for complex queries.}''
{\tool} then provides suggestions to fix the misalignments, which further reduces error correction efforts.
P8 wrote, ``\textit{It is helpful to edit by simply clicking buttons.}''
These two features effectively incorporate human knowledge into the system, providing a straightforward way for humans to collaborate with the LLM.
Fourth, {\tool} reduces mental effort in understanding complex database schemas and dataset composition through visualization.
The visualization helps improve human control throughout the annotation.
P2 wrote, ``\textit{I really like the schema page, I think it helps visualize the structure of the table a lot better than seeing it in a JSON format.}''
P6 wrote, ``\textit{The data analysis helps me understand the data and where I should go on.}''

While {\tool} is designed for text-to-SQL, these four design ideas can be applied to other relevant domains.
\edit{For example, in semantic parsing for mathematical expressions, the system could suggest formulas based on structural patterns (e.g., equations with specific operators or the number of variables) and allow humans to verify mathematical relationships while visualizing expression trees. The system could detect inconsistencies between the natural language description and mathematical notation, suggesting corrections based on common mathematical writing conventions.}
\edit{In semantic parsing for other formal languages like Python, the system could suggest code snippets in a configurable manner by code properties, such as test cases to indicate semantic patterns and control flow graphs to indicate syntactic patterns.}
\edit{In image captioning, the system could generate structured captions following predefined templates like ``A [subject] is [action] near a [object]''. The system would render the scene using predefined objects. This visualization helps humans quickly verify if the caption accurately describes the key objects and their spatial relationships in the image.}

\edit{{\tool} also ensures fairness and reduces bias in data annotation by providing features such as confidence scoring and diversity checking. These features enable users to identify and correct biases introduced by AI systems, keeping the annotation process transparent and under human control.}

\subsection{Limitation and Future Work}
While {\tool} shows significant effectiveness in text-to-SQL data annotation, participants suggested several improvements for future work.
P1 and P7 pointed out the suggested NL question may be inaccurate when the SQL query is complex, which requires a more advanced SQL-to-text approach in the backend.
P7 mentioned that generating the suggested NL question can take seconds and hopes for optimization of this generation time.
We believe this can be optimized by generating subsequent SQL queries and NL questions in the backend while users process previous data. Users will not have to wait for the system to dynamically generate suggested data.
P3 and P11 suggested that adding a search function for entity names within complex schema visualizations would be helpful.
To further increase controllability, P4 suggested adding options to control the properties in generated SQL rather than making it completely random.
We think all of these suggestions are valuable for future improvements in {\tool}.

\section{Conclusion}
This paper presents {\tool}, a novel interactive text-to-SQL annotation system that enables users to create high-quality, schema-specific text-to-SQL datasets.
{\tool} integrates multiple functionalities, including schema customization, database synthesis, query alignment, dataset analysis, and additional features such as confidence scoring.
A user study with 12 participants demonstrates that by combining these features, {\tool} significantly reduces annotation time while improving the quality of the annotated data.
{\tool} effectively bridges the gap resulting from insufficient training and evaluation datasets for new or unexplored schemas.

\begin{acks}
We thank the anonymous reviewers for their helpful and detailed feedback, as well as the time and care they dedicated to reviewing our work. We also express our gratitude to all the participants in the interview and user study for their valuable comments. This work was supported by Adobe during the first author's internship.
\end{acks}

\bibliographystyle{ACM-Reference-Format}
\bibliography{reference}

\appendix

\section{Additional User Study: Database Customization}
\label{app:additional_study}

The text-to-SQL data annotation process can be divided into two stages: (1) database schema customization and (2) text-to-SQL data annotation based on a provided schema.
As the core contribution of {\tool}, we primarily focused on evaluating the text-to-SQL data annotation stage in the main study. 
In this section, we reported the evaluation of database schema customization as an additional study.
The participants and conditions were the same as the main study.

For the results, we began by analyzing participants' schema customization performance across different conditions, focusing on both accuracy and speed. 
Then, we reported user perceptions of the various conditions, including their confidence and cognitive load.

\subsection{Tasks}
To assess schema customization performance, we created a pool of schema editing tasks. For each sampled schema, we manually created 30 tasks requiring edits over the existing schema, e.g., ``\textit{Add a new column 'Founded' (date) to the 'airlines' table.}''
Participants were expected to complete these tasks sequentially, as some tasks depended on the completion of previous ones. 
We maintained a consistent distribution of task types (e.g., the number of ``add column'' tasks) across different schemas.

\subsection{Protocol}
There were three sessions corresponding to three conditions.
For each session, participants were given a database schema and 30 tasks describing how to edit the schema. The three sessions corresponded to three database schemas. Given the excessive number of tasks, participants were asked to complete as many as possible within the 5-minute time limit.

After each session, participants completed a post-task survey to rate their experience with the assigned condition. The survey included the System Usability Score (SUS)~\cite{sus} and NASA Task Load Index (TLX)~\cite{NASA-TLX} questionnaires, using 7-point Likert scales to assess their perceptions.

\subsection{Schema Customization Performance}
We manually review the correctness of participants' completed tasks. For each task, we mark it as correct if the participant met all requirements; otherwise, we mark it as wrong. Table~\ref{tab:schema_edit_performance} reports the number of completed tasks and their accuracy for each condition.

The results show that using {\tool} helps users achieve the highest customization speed (10.73) and accuracy (96.86\%) within 5 minutes. When manually customizing the schema in JSON file, users achieve the lowest customization speed (7.18) and accuracy (81.27\%). When using ChatGPT, users achieve intermediate speed (9.73) and accuracy (67.48\%).

When using ChatGPT, we observe that some participants directly copy and paste the entire schema and task descriptions into the prompt. This approach works when the schema is simple and includes only a few tables and columns. However, as the schema becomes more complex, ChatGPT may make mistakes due to hallucination~\cite{llm_hallucination}.

When using {\tool}, while there is no LLMs in the backend, participants complete schema customization tasks very quickly by directly editing the graph. Users can make precise edits and immediately detect errors through the visualization. In fact, except for one participant who misunderstand entity names, all other participants achieve 100\% task completion accuracy.

\begin{table}[htb]
    \centering
    \caption{Schema Customization Speed and Accuracy.}
    \vspace{-2.5mm}
    \resizebox{\linewidth}{!}{%
    \begin{tabular}{lccc}
        \toprule
              & \textbf{Completed Tasks} & \textbf{Correct Tasks} & \textbf{Accuracy} \\
        \midrule
        Manual  & 7.18  & 5.82 & 81.27\% \\
        AI-assistant   & 9.73   & 6.64 & 67.48\% \\
        {\tool} & \textbf{10.73}   & \textbf{10.45} & \textbf{96.86\%} \\
        \bottomrule
    \end{tabular}
    }
    
    \label{tab:schema_edit_performance}
\end{table}

\subsection{User Cognitive Load \& Usability Rating}

\begin{figure*}[ht]
  \centering
  \includegraphics[width=\textwidth]{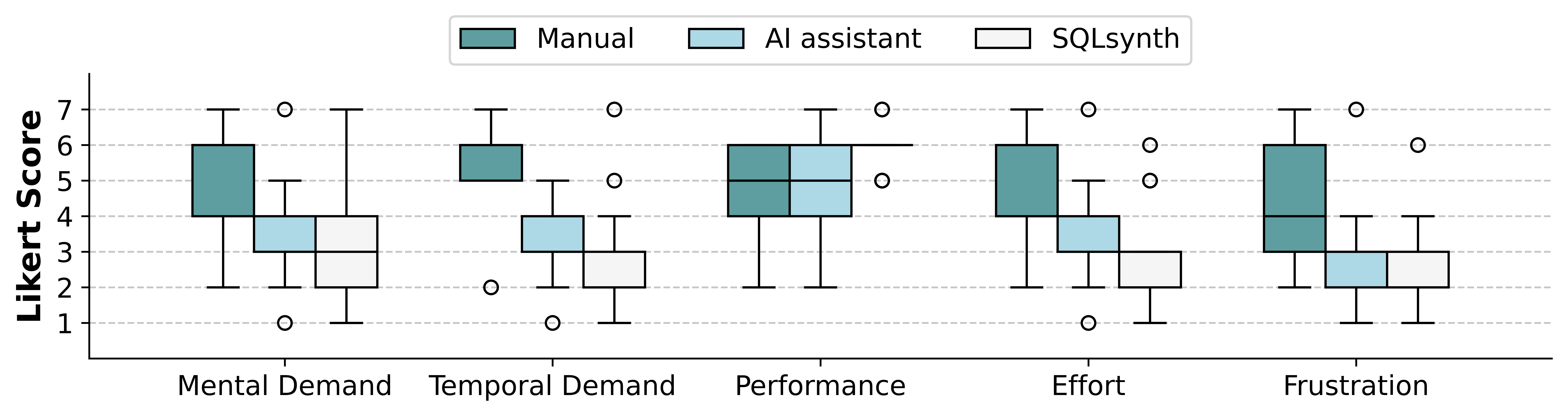}
  \caption{NASA Task Load Index Ratings of Schema Customization}
  \label{fig:cognitive2}
\end{figure*}

Figure~\ref{fig:cognitive2} illustrates participants' ratings on the five cognitive load factors from the NASA TLX questionnaire. The ANOVA test reveals statistically significant ($\alpha = 0.05$) differences in means for all factors: Mental Demand ($p = 2.95 \times 10^{-3}$), Temporal Demand ($p = 3.31 \times 10^{-4}$), Performance ($p = 4.76 \times 10^{-2}$), Effort ($p = 1.81 \times 10^{-2}$), and Frustration ($p = 3.49 \times 10^{-2}$).

Participants appreciate visualizing the schema as an editable graph, which facilitates both comprehension and modification.
P2 report, ``\textit{I really like the schema editing page, I think it helps visualize the structure of the table a lot better than seeing it in a JSON format.}''
P11 write, ``\textit{Schema editing was easier and the ability to visualize the edits was very useful.}''
P9 comment, ``\textit{I definitely found the tool much more handy to make quick edits and understand the dependencies between different tables.}''


\begin{figure*}[ht]
  \centering
  \includegraphics[width=0.8\textwidth]{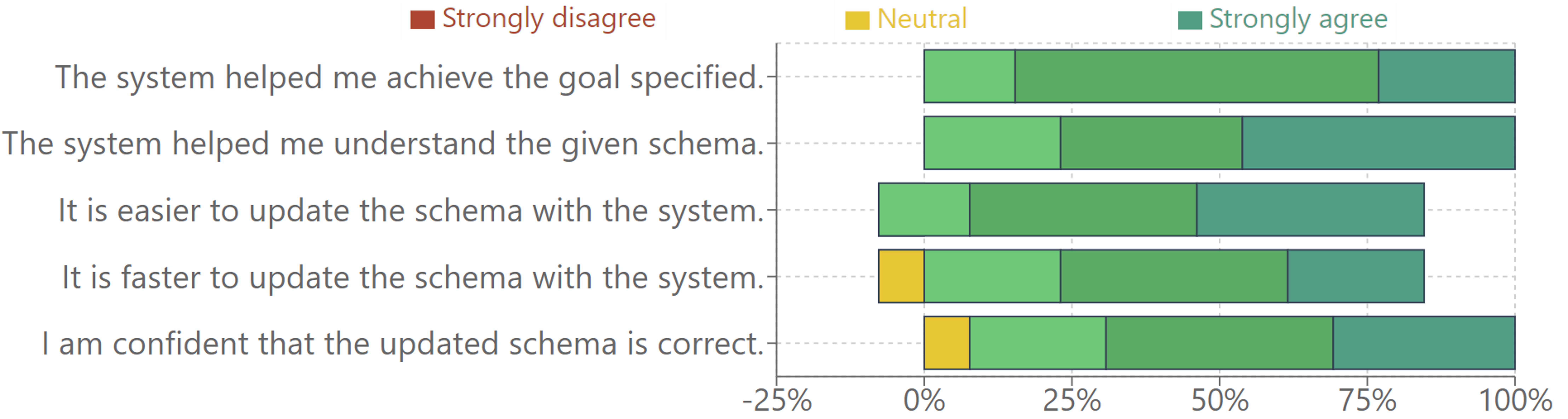}
  \caption{SUS Scores of Schema Customization}
  \label{fig:sus2}
\end{figure*}

Figure~\ref{fig:sus2} presents SUS scores for Task 2, revealing highly positive user perceptions across all dimensions. 
Participants agree or strongly agree that {\tool} help them understand schemas. 
They also find updating schemas easier and faster, with high confidence in the results.

\begin{table*}[!h]
\centering
\caption{PCFG Grammar for SQL Query Generation}
\label{tab:sql-pcfg}
\begin{tabular}{l@{\ }c@{\ }l}
\toprule
Non-terminal && Production Rule (Probability) \\
\midrule
$\langle$Query$\rangle$ & $\rightarrow$ & $\langle$SelectClause$\rangle$ $\langle$FromClause$\rangle$ [$\langle$WhereClause$\rangle$] [$\langle$GroupByClause$\rangle$] [$\langle$OrderByClause$\rangle$]  (1.0) \\[1ex]
$\langle$SelectClause$\rangle$ & $\rightarrow$ & \texttt{SELECT} $\langle$ColumnList$\rangle$  (0.9) \\
& $|$ & \texttt{SELECT DISTINCT} $\langle$ColumnList$\rangle$  (0.1) \\[1ex]
$\langle$ColumnList$\rangle$ & $\rightarrow$ & $\langle$Column$\rangle$  (0.6) \\
& $|$ & $\langle$Column$\rangle$\texttt{,} $\langle$ColumnList$\rangle$  (0.4) \\[1ex]
$\langle$Column$\rangle$ & $\rightarrow$ & $\langle$TableName$\rangle$\texttt{.}$\langle$ColumnName$\rangle$  (0.7) \\
& $|$ & $\langle$AggregateFunction$\rangle$\texttt{(}$\langle$TableName$\rangle$\texttt{.}$\langle$ColumnName$\rangle$\texttt{)}  (0.3) \\[1ex]
$\langle$FromClause$\rangle$ & $\rightarrow$ & \texttt{FROM} $\langle$TableName$\rangle$  (0.4) \\
& $|$ & \texttt{FROM} $\langle$TableName$\rangle$ $\langle$JoinClause$\rangle$  (0.6) \\[1ex]
$\langle$JoinClause$\rangle$ & $\rightarrow$ & $\langle$JoinType$\rangle$ \texttt{JOIN} $\langle$TableName$\rangle$ \texttt{ON} $\langle$JoinCondition$\rangle$  (0.7) \\
& $|$ & $\langle$JoinType$\rangle$ \texttt{JOIN} $\langle$TableName$\rangle$ \texttt{ON} $\langle$JoinCondition$\rangle$ $\langle$JoinClause$\rangle$  (0.3) \\[1ex]
$\langle$JoinType$\rangle$ & $\rightarrow$ & \texttt{INNER}  (0.4) \\
& $|$ & \texttt{LEFT}  (0.3) \\
& $|$ & \texttt{RIGHT}  (0.2) \\
& $|$ & \texttt{FULL}  (0.1) \\[1ex]
$\langle$JoinCondition$\rangle$ & $\rightarrow$ & $\langle$Column$\rangle$ \texttt{=} $\langle$Column$\rangle$  (0.8) \\
& $|$ & $\langle$JoinCondition$\rangle$ \texttt{AND} $\langle$JoinCondition$\rangle$  (0.2) \\[1ex]
$\langle$WhereClause$\rangle$ & $\rightarrow$ & \texttt{WHERE} $\langle$Condition$\rangle$  (1.0) \\[1ex]
$\langle$Condition$\rangle$ & $\rightarrow$ & $\langle$Column$\rangle$ $\langle$Operator$\rangle$ $\langle$Value$\rangle$  (0.5) \\
& $|$ & $\langle$Condition$\rangle$ \texttt{AND} $\langle$Condition$\rangle$  (0.3) \\
& $|$ & $\langle$Condition$\rangle$ \texttt{OR} $\langle$Condition$\rangle$  (0.2) \\[1ex]
$\langle$GroupByClause$\rangle$ & $\rightarrow$ & \texttt{GROUP BY} $\langle$ColumnList$\rangle$  (1.0) \\[1ex]
$\langle$OrderByClause$\rangle$ & $\rightarrow$ & \texttt{ORDER BY} $\langle$ColumnList$\rangle$ [\texttt{ASC} $|$ \texttt{DESC}]  (1.0) \\[1ex]
$\langle$AggregateFunction$\rangle$ & $\rightarrow$ & \texttt{COUNT}  (0.3) \\
& $|$ & \texttt{SUM}  (0.2) \\
& $|$ & \texttt{AVG}  (0.2) \\
& $|$ & \texttt{MIN}  (0.15) \\
& $|$ & \texttt{MAX}  (0.15) \\[1ex]
$\langle$Operator$\rangle$ & $\rightarrow$ & \texttt{=}  (0.3) \\
& $|$ & \texttt{<}  (0.1) \\
& $|$ & \texttt{>}  (0.1) \\
& $|$ & \texttt{<=}  (0.1) \\
& $|$ & \texttt{>=}  (0.1) \\
& $|$ & \texttt{<>}  (0.1) \\
& $|$ & \texttt{LIKE}  (0.1) \\
& $|$ & \texttt{IN}  (0.1) \\[1ex]
$\langle$Value$\rangle$ & $\rightarrow$ & $\langle$Number$\rangle$  (0.4) \\
& $|$ & $\langle$String$\rangle$  (0.4) \\
& $|$ & $\langle$Column$\rangle$  (0.2) \\[1ex]
\multicolumn{3}{l}{$\langle$TableName$\rangle$, $\langle$ColumnName$\rangle$, $\langle$Number$\rangle$, $\langle$String$\rangle$ represent database-specific values or literals} \\
\bottomrule
\end{tabular}
\end{table*}

\section{Example of PCFG in SQL Sampling}
\label{app:pcfg}

Table~\ref{tab:sql-pcfg} exemplifies the Probabilistic Context-Free Grammar (PCFG) used for random sampling in Section~\ref{sec:sql_sampling}.
Note that these probabilities are illustrative examples and not derived from real-world query distributions. In practice, these probabilities can be tuned by a large collection of real-world SQL queries. To protect customer data privacy, we do not display the actual probability distribution. 

The grammar begins with the $\langle$Query$\rangle$ non-terminal, which expands into the fundamental components of an SQL query, including $\langle$SelectClause$\rangle$, $\langle$FromClause$\rangle$, and optional $\langle$JoinClause$\rangle$, $\langle$WhereClause$\rangle$, $\langle$GroupByClause$\rangle$, and $\langle$OrderByClause$\rangle$. Each clause is further defined by its constituent parts. For instance, the $\langle$SelectClause$\rangle$ can be either a simple \texttt{SELECT} or \texttt{SELECT DISTINCT}, followed by a $\langle$ColumnList$\rangle$. The $\langle$ColumnList$\rangle$ can be a single $\langle$Column$\rangle$ or recursively defined as $\langle$Column$\rangle$\texttt{,} $\langle$ColumnList$\rangle$, allowing for multiple column selections. Columns can be either simple references ($\langle$TableName$\rangle$\texttt{.}$\langle$ColumnName$\rangle$) or aggregations ($\langle$AggregateFunction$\rangle$ \texttt{(}$\langle$TableName$\rangle$\texttt{.}$\langle$ColumnName$\rangle$\texttt{)}). The grammar also defines common SQL operations and functions, such as comparison operators (\texttt{=}, \texttt{<}, \texttt{>}, etc.) and aggregate functions (\texttt{COUNT}, \texttt{SUM}, \texttt{AVG}, etc.). 
Probabilities associated with each production rule guide the random sampling process. 
For example, a $\langle$SelectClause$\rangle$ has a 90\% chance of being a simple \texttt{SELECT} and a 10\% chance of being \texttt{SELECT DISTINCT} in our example. 

\section{Prompts Used in {\tool}}
\label{app:prompt}

\edit{In this section, we present the prompts used in {\tool} for different functionalities.Each prompt is carefully designed to guide the language model in generating specific outputs while maintaining consistency and quality. The prompt includes relevant context, objective, and structured output format for the following parsing.}

\begin{figure*}[ht]
  \centering
\includegraphics[width=\textwidth]{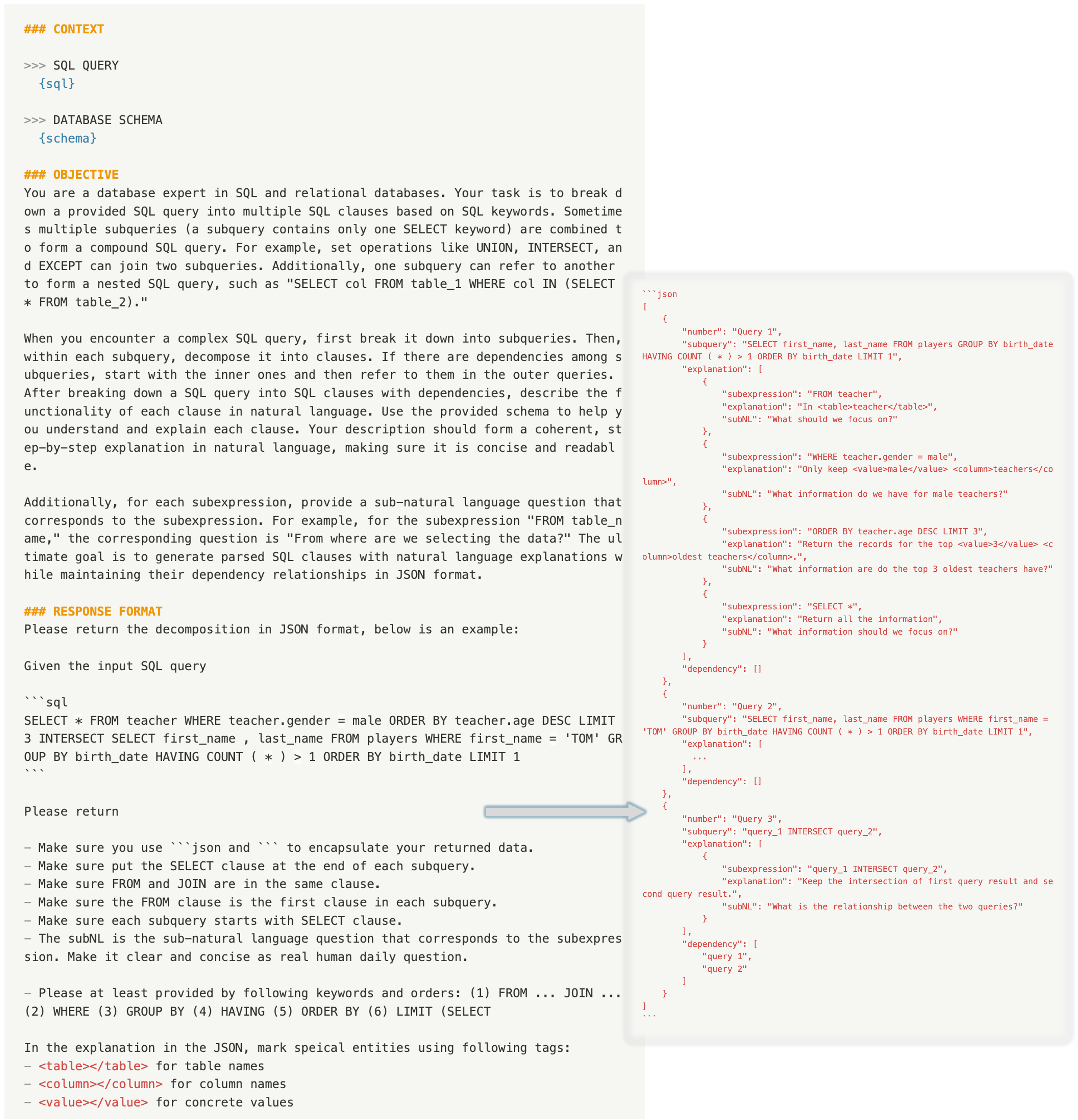}
  \caption{Prompt of Step-by-step Explanation \& Sub-Question Generation.}
  \label{fig:prompt_explanation}
\end{figure*}

\begin{figure*}[ht]
  \centering
\includegraphics[width=0.65\textwidth]{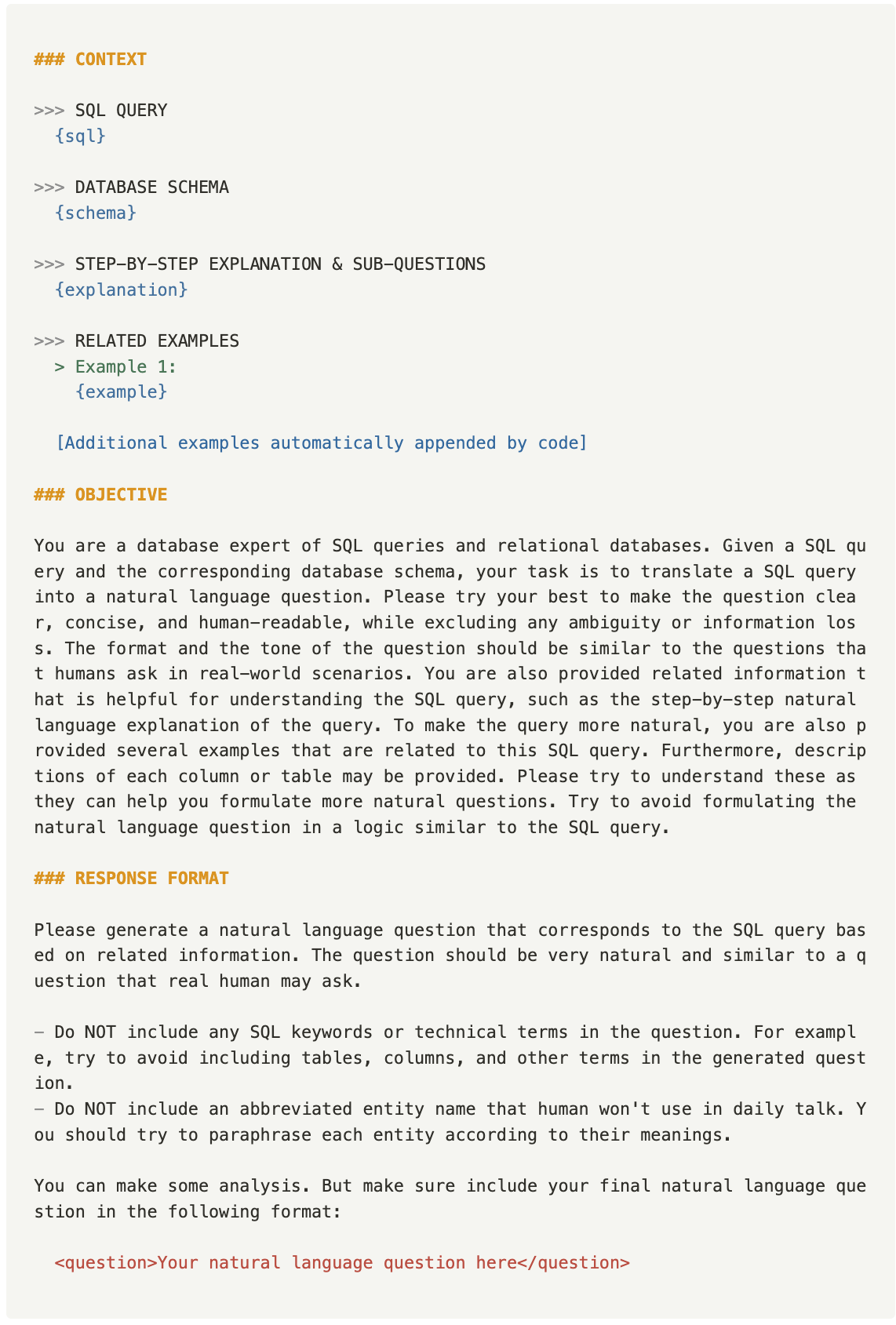}
  \caption{Prompt of Natural Language Question Generation.}
  \label{fig:prompt_question}
\end{figure*}

\begin{figure*}[ht]
  \centering
\includegraphics[width=0.65\textwidth]{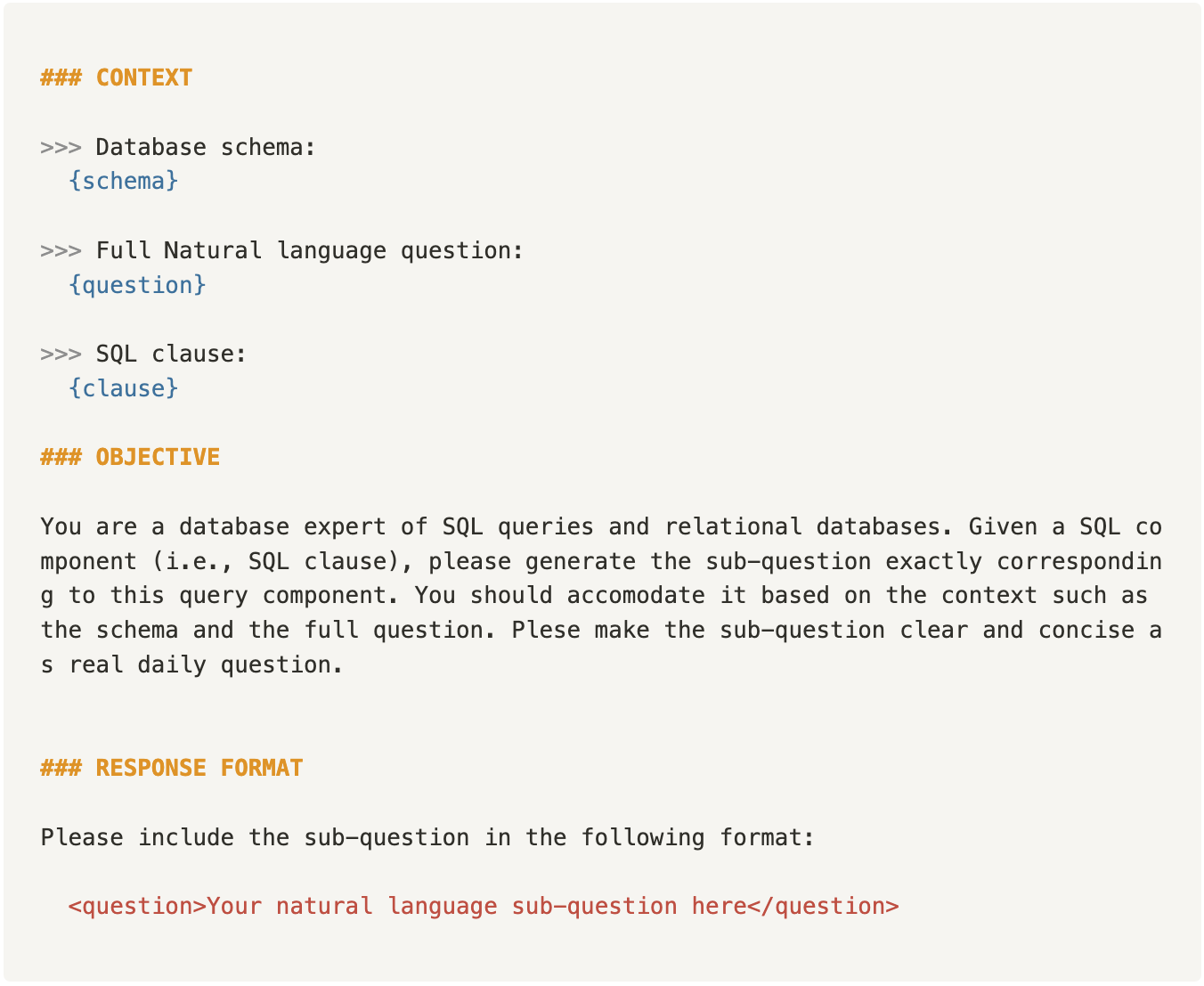}
  \caption{Prompt of Sub-question Generation.}
  \label{fig:prompt_subquestion}
\end{figure*}

\begin{figure*}[ht]
  \centering
\includegraphics[width=0.65\textwidth]{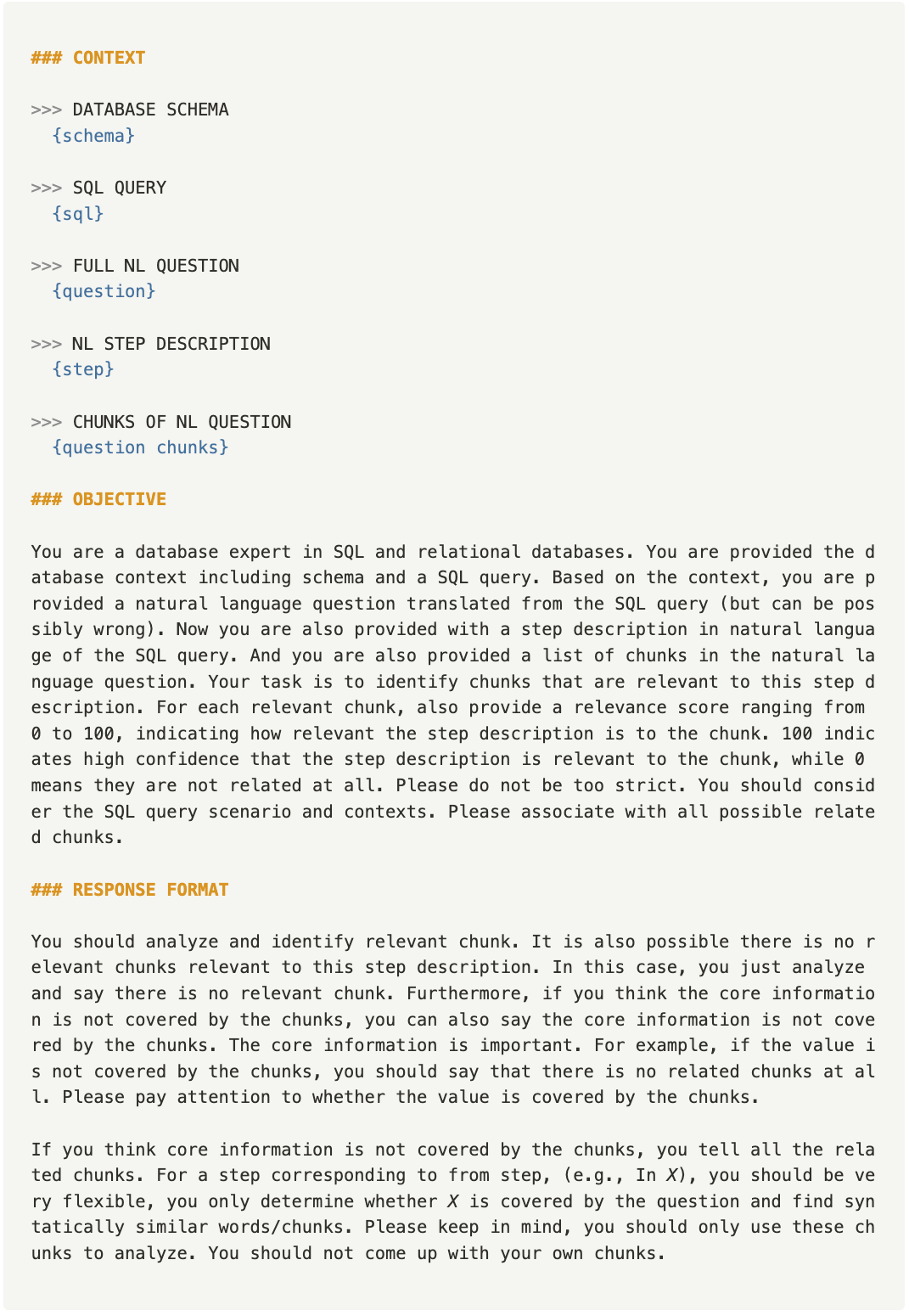}
  \caption{Prompt of Alignment Analysis Generation}
  \label{fig:prompt_alignment_analysis}
\end{figure*}

\begin{figure*}[ht]
  \centering
\includegraphics[width=0.65\textwidth]{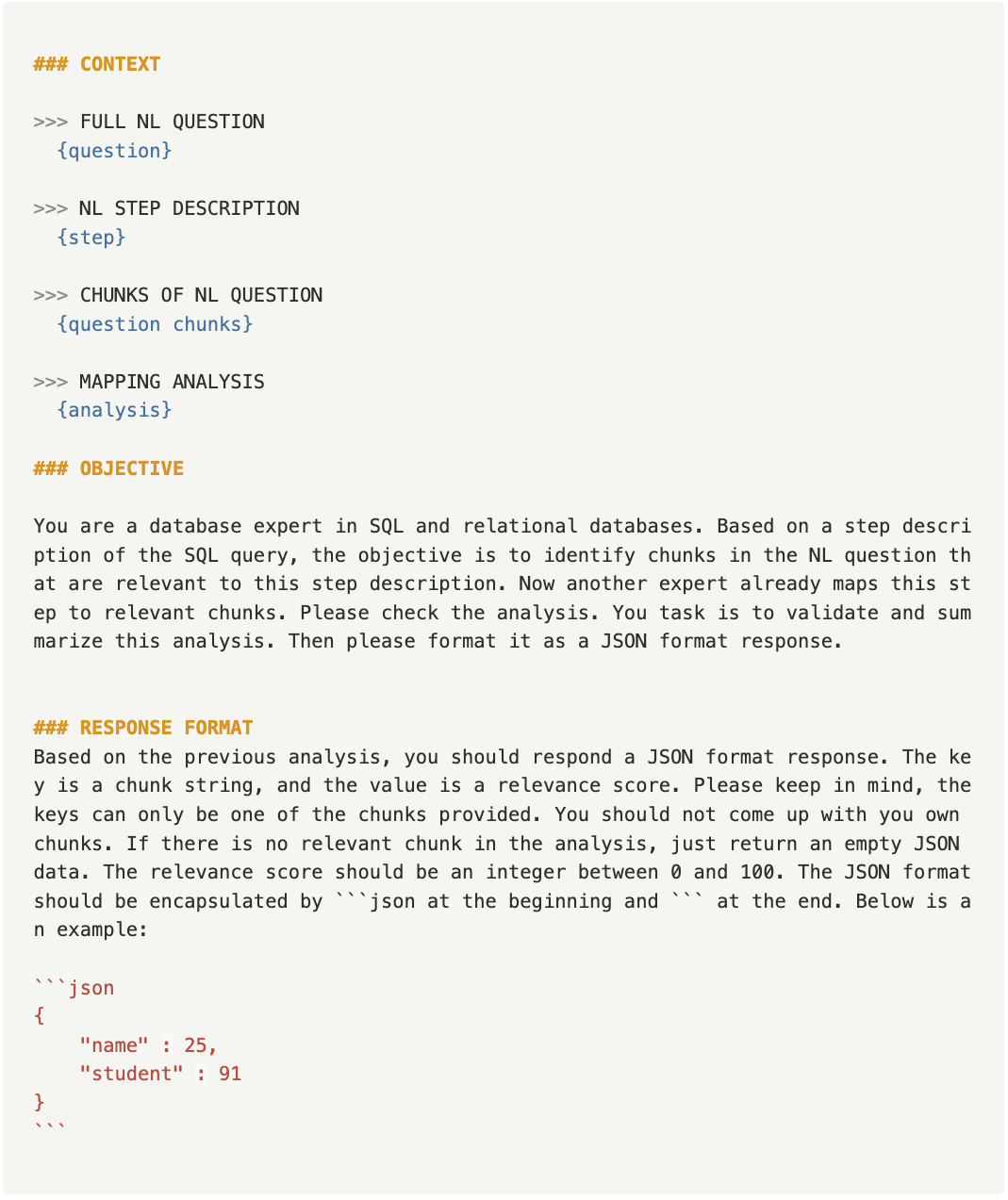}
  \caption{Prompt of Alignment Mapping Generation}
  \label{fig:prompt_alignment_map}
\end{figure*}

\begin{figure*}[ht]
  \centering
\includegraphics[width=0.65\textwidth]{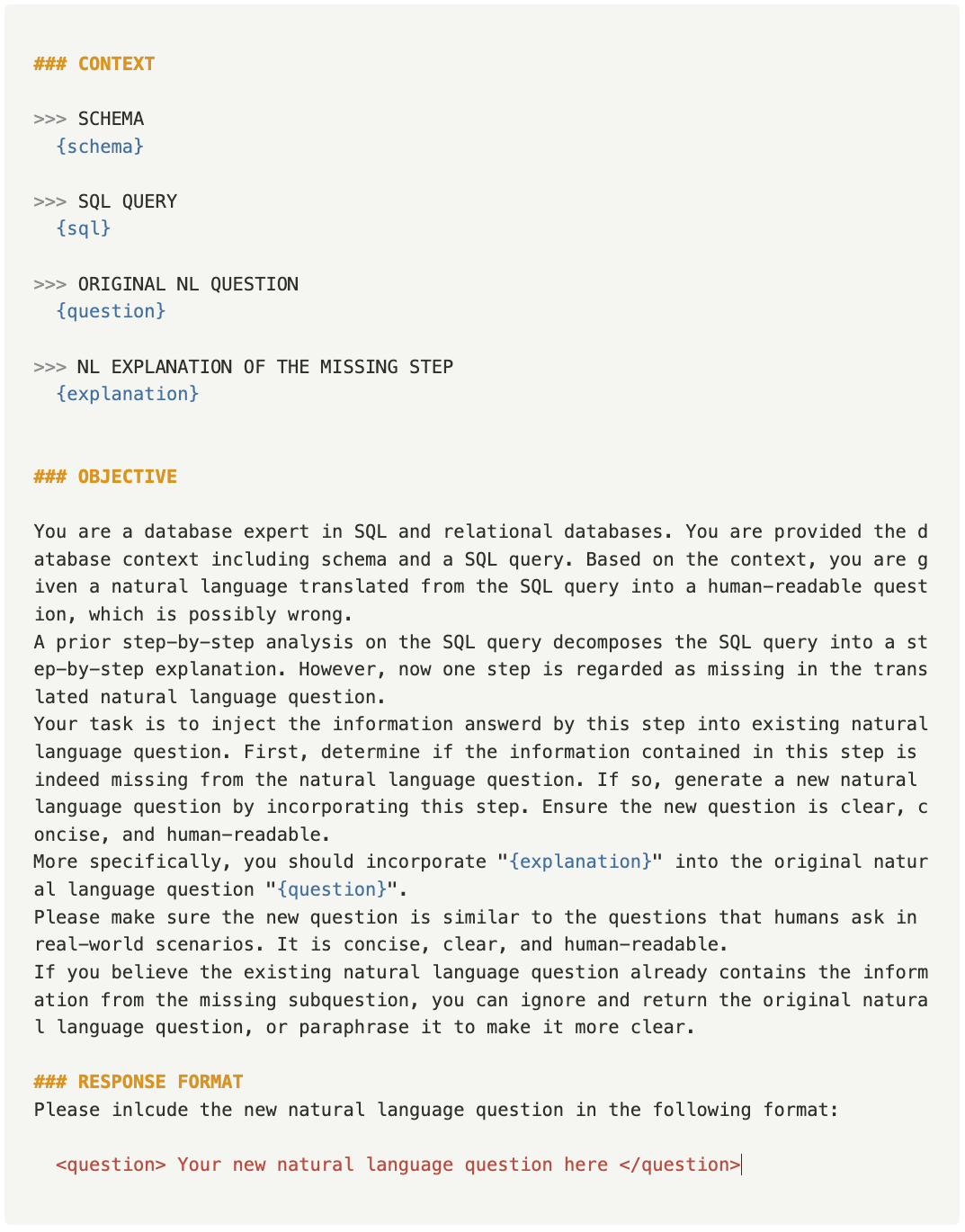}
  \caption{Prompt of Updating NL question by emphasizing a certain step}
  \label{fig:prompt_inject}
\end{figure*}

\begin{figure*}[ht]
  \centering
\includegraphics[width=0.65\textwidth]{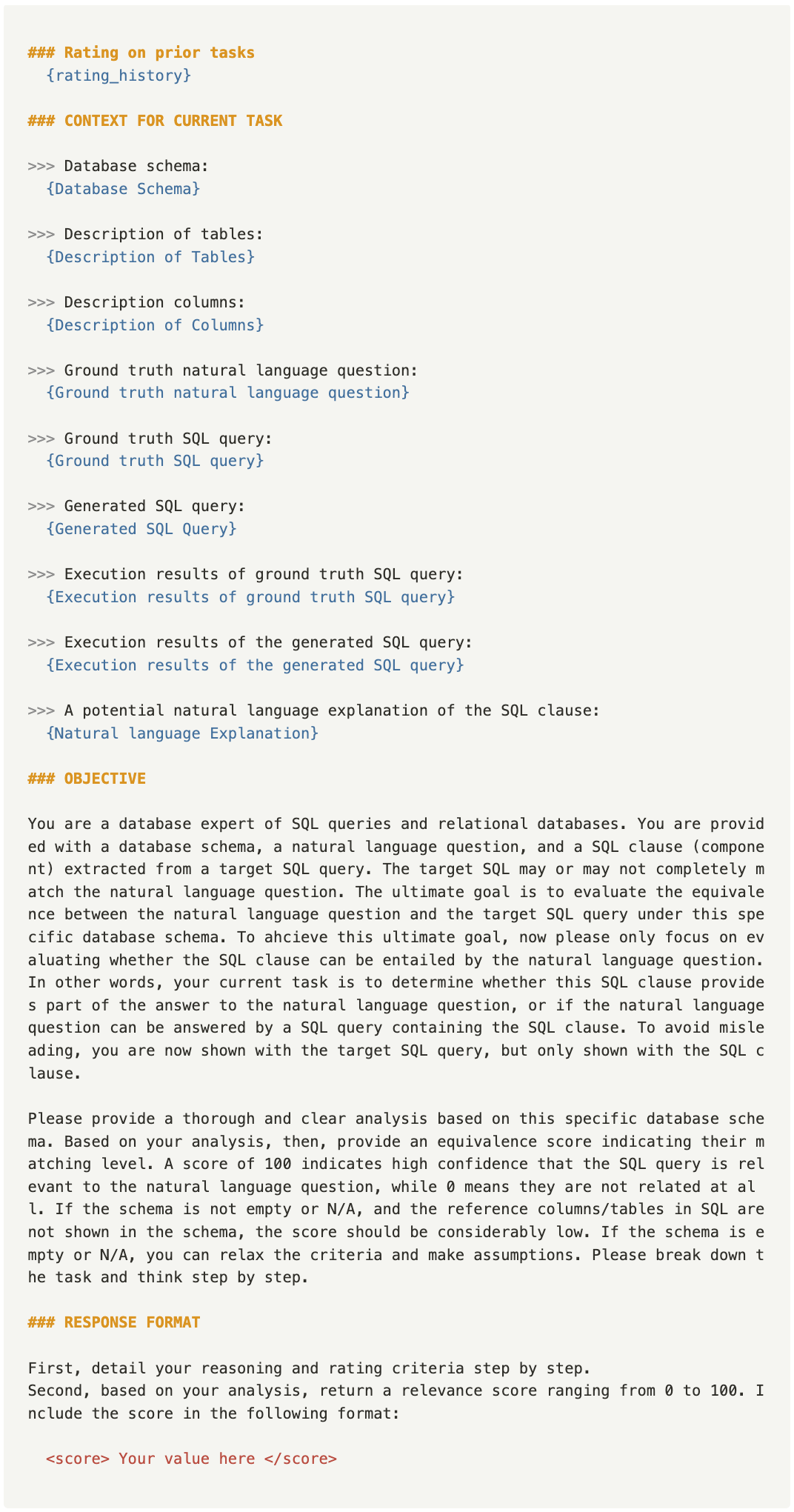}
  \caption{Prompt of NL-SQL Equivalence Analysis and Scoring}
  \label{fig:prompt_equivalence}
\end{figure*}

\edit{Figure~\ref{fig:prompt_explanation} shows the prompt for generating step-by-step explanations and sub-questions. Note that this prompt is invoked and the LLM is called only when the rule-based explanation generation fails. Figure~\ref{fig:prompt_question} displays the prompt for natural language question generation. Figure~\ref{fig:prompt_subquestion} illustrates the prompt for sub-question generation. Figures~\ref{fig:prompt_alignment_analysis} and~\ref{fig:prompt_alignment_map} show the prompts for alignment analysis and mapping generation, respectively. 
As a core functionality in {\tool}, we decompose this task into two smaller subtasks: free-form text analysis and structured output generation. This decomposition is motivated by research~\cite{multi_agent_collaboration} showing that multi-agent collaboration leads to improved generation accuracy.
Figure~\ref{fig:prompt_inject} presents the prompt for updating natural language questions by emphasizing specific steps. 
Finally, Figure~\ref{fig:prompt_equivalence} shows the prompt for analyzing and scoring NL-SQL equivalence.}

\vspace*{\fill}
\vspace*{\fill}

\vspace*{\fill}
\vspace*{\fill}

\vspace*{\fill}
\vspace*{\fill}

\vspace*{\fill}
\vspace*{\fill}

\vspace*{\fill}
\vspace*{\fill}

\vspace*{\fill}
\vspace*{\fill}

\vspace*{\fill}
\vspace*{\fill}

\end{document}